\pdfoutput=1
\documentclass[%
  aps,
  prb,
  showpacs,
  citeautoscript,
  superscriptaddress,
  amsmath,
  amssymb,
  reprint
]{revtex4-1}

\usepackage[utf8]{inputenc}
\usepackage[T1]{fontenc}
\usepackage[colorlinks=true,urlcolor=blue,linkcolor=blue,citecolor=blue]{hyperref}

\usepackage{graphicx}
\usepackage{color}
\usepackage[version=3]{mhchem}
\usepackage{units}

\DeclareMathAlphabet{\mathcal}{OMS}{cmsy}{m}{n}
\graphicspath{{./}{figs/}}
\begin{document}

\author{Tuomas P.\ Rossi}
\affiliation{Department of Physics, Chalmers University of Technology, 412 96 Gothenburg, Sweden}
\author{Timur Shegai}
\affiliation{Department of Physics, Chalmers University of Technology, 412 96 Gothenburg, Sweden}
\author{Paul Erhart}
\affiliation{Department of Physics, Chalmers University of Technology, 412 96 Gothenburg, Sweden}
\author{Tomasz J.\ Antosiewicz}
\email{tomasz.antosiewicz@fuw.edu.pl}
\affiliation{Faculty of Physics, University of Warsaw, Pasteura 5, 02-093 Warsaw, Poland}
\affiliation{Department of Physics, Chalmers University of Technology, 412 96 Gothenburg, Sweden}

\title{Strong plasmon-molecule coupling at the nanoscale\\ revealed by first-principles modeling}

\date{\today}

\begin{abstract}
Strong light-matter interactions in both the single-emitter and collective strong coupling regimes attract significant attention due to emerging quantum and nonlinear optics applications, as well as opportunities for modifying material-related properties.
Further exploration of these phenomena requires an appropriate theoretical methodology, which is demanding since polaritons are at the intersection between quantum optics, solid state physics and quantum chemistry. Fortunately, however, nanoscale polaritons can be realized in small plasmon-molecule systems, which in principle allows treating them using \emph{ab initio} methods, although this has not been demonstrated to date.
Here, we show that time-dependent density-functional theory (TDDFT) calculations can access the physics of nanoscale plasmon-molecule hybrids and predict vacuum Rabi splitting in a system comprising a few-hundred-atom aluminum nanoparticle interacting with one or several benzene molecules.
We show that the cavity quantum electrodynamics approach holds down to resonators on the order of a few cubic nanometers, yielding a single-molecule coupling strength exceeding \unit[200]{meV} due to a massive vacuum field value of $E_\mathrm{vac} = \unit[4.5]{V/nm}$.
In a broader perspective, our approach enables parameter-free in-depth studies of polaritonic systems, including ground state, chemical and thermodynamic modifications of the molecules in the strong-coupling regime, which may find important use in emerging applications such as cavity enhanced catalysis.
\end{abstract}

\maketitle

Light-matter interaction between an optical mode and a quantum emitter is accurately described in terms of the cavity quantum electrodynamics (cQED) formalism \cite{scully_zubairy_1997}.
The interaction can be weak or strong, depending on circumstances, leading to drastically different behavior.
The regime of strong light-matter coupling, unlike its weak counterpart, leads to formation of hybrid cavity-emitter eigenmodes manifested in coherent energy exchange between the system subsystems occurring on timescales that are (much) faster than the corresponding damping rates.
Thus, the emitter and the cavity form a unified light-matter hybrid polariton, whose properties, including spontaneous emission and chemical potential, can be tuned \cite{RepProgPhys_78_013901_torma, 2018_ASCP_5_24_baranov, NatRevPhys_1_19_frisk}.
Because of their compositional nature, polaritons are useful for photon-photon interactions, which leads to remarkable nonlinear and quantum optical phenomena, including photon blockade and Bose-Einstein condensation of exciton polaritons \cite{Nature_432_200_yoshie, Nature_436_87_birnbaum, Nature_443_409_kasprzak}.
On the other hand, strong light-matter coupling may lead to changes in emitter properties, including photochemical rates \cite{AngChem_51_1592_ebbesen, PRL_116_238301_herrera, NComm_7_13841_galego, ACSP_5_167_martinez, SciAdv_4_eaas9552_munkhbat} and exciton transport \cite{PRL_114_196402_feist, AngChem_55_6202_zhong}.

Traditionally cavity-exciton polaritons were realized in atomic or solid state systems utilizing high quality factor optical cavities at low temperatures.
These efforts resulted in remarkable effects such as polariton Bose-Einstein condensates and superfluidity \cite{NatPhys_2_81_khitrova}.
Polaritonic behavior associated with ``dressing'' of the emitter by a cavity field is usually captured by traditional quantum optical approaches such as Jaynes-Cummings or Dicke models \cite{ProcIEEE_51_89_haynes, PhilTransA_369_1137_garraway, RepProgPhys_78_013901_torma, 2018_ASCP_5_24_baranov, NatRevPhys_1_19_frisk}.
However, these quantum optical formalisms treat matter in an extremely simplified manner, that is, as a two-level system, leading to oversimplifications and inconsistencies in the description of material subpart.

More advanced theoretical techniques developed recently allow for more sophisticated effects including, e.g., multiple electronic resonances, accounting for atomic vibrations, and light-matter interactions beyond the point dipole approximation.
Significant progress along these lines has been achieved by several groups using various quantum optical and quantum chemistry methods \cite{PRL_116_238301_herrera, NComm_7_13841_galego, ACSP_5_167_martinez, PRX_5_041022_galego, PRA_90_012508_ruggenthaler, JCTC_13_4324_ling, NL_18_2358_neuman}.
However, typically either molecules or electromagnetic fields in these approaches are treated in a simplified manner.
Here, we model the entire plasmon-exciton system by time-dependent density-functional theory (TDDFT) \cite{1984_PRL_52_997_runge}.

Plasmon-molecule interactions can be modeled computationally using density-functional theory (DFT) \cite{1964_PhysRev_136_B864_hohenberg, 1965_PhysRev_140_A1133_kohn} and/or TDDFT approaches, thanks to the relatively small number of atoms involved in these interactions.
TDDFT allows the whole system to be treated on the same footing, enabling one to track effects related to modification of the matter subpart, which are inaccessible by purely quantum optical or classical electromagnetism methods.
In fact, analytical and numerical quantum chemistry approaches, including DFT and TDDFT, have been successfully applied to study plasmon-molecule interactions, charge transfer, chemical enhancement and electromagnetic effects in surface-enhanced Raman scattering (SERS) experiments \cite{1996_JPhysChem_100_3199_arenas, 2002_JChemPhys_116_7207_arenas, 2004_JPhysChemA_108_4187_bjerneld, 2006_JACS_128_2911_zhao, 2008_ChemSocRev_37_1061_jensen, 2011_JChemPhys_135_1344103_morton, 2001_ChemPhysLett_342_135_corni, 2008_JPCC_112_5605_lombardi, 2009_JACS_131_14390_shegai}.
However, vacuum Rabi splitting and strong plasmon-molecule coupling have not been in the focus.

Related to the above, quantum electrodynamics density-functional theory (QED-DFT) has been developed recently \cite{PRA_90_012508_ruggenthaler}, enabling the modeling of strong coupling between electromagnetic cavity modes and electronic excitations \cite{2017_PNAS_114_3026_flick, 2018_Nanophotonics_7_1479_flick}.
While the commonly-used TDDFT approaches with classical description of electric fields may not be able to describe the full range of quantum optical phenomena that QED-DFT aims at, we demonstrate in this paper that the cavity mode created by a localized surface plasmon resonance as well as its strong coupling with excitons can be described already within the standard TDDFT.

Additionally, many recent experimental observations cannot be easily explained with current theories \cite{2013_AdvMater_25_2481_hutchison, 2015_NatureMater_14_1123_orgiu, 2016_AngChemIE_55_11462_thomas}.
Here, we take a step towards helping to elucidate such observations by demonstrating the possibilities of TDDFT for studying polariton physics.
We argue that theoretical predictions obtained on this small scale may be useful for understanding and modeling of more extended systems.

We model light-matter interactions by employing the real-time-propagation (RT) TDDFT approach \cite{1996_PRB_54_4484_yabana} based on the localized basis sets \cite{2015_PRB_91_115431_kuisma, 2009_PRB_80_195112_larsen} as implemented in the open-source GPAW package \cite{2005_PRB_71_035109_mortensen, 2010_JPhysCondMat_22_253202_enkovaara}.
This RT-TDDFT code is combined with extensive analysis tools \cite{2017_JCTC_13_4779_rossi}, that are utilized for analyzing the electron-hole transition contributions to resonances and visualizing them as transition contribution maps (TCM) \cite{2013_ACSNano_7_10263_malola, 2017_JCTC_13_4779_rossi}.
See Methods for detailed description.

\begin{figure}
\centering
\includegraphics[width=8.5cm]{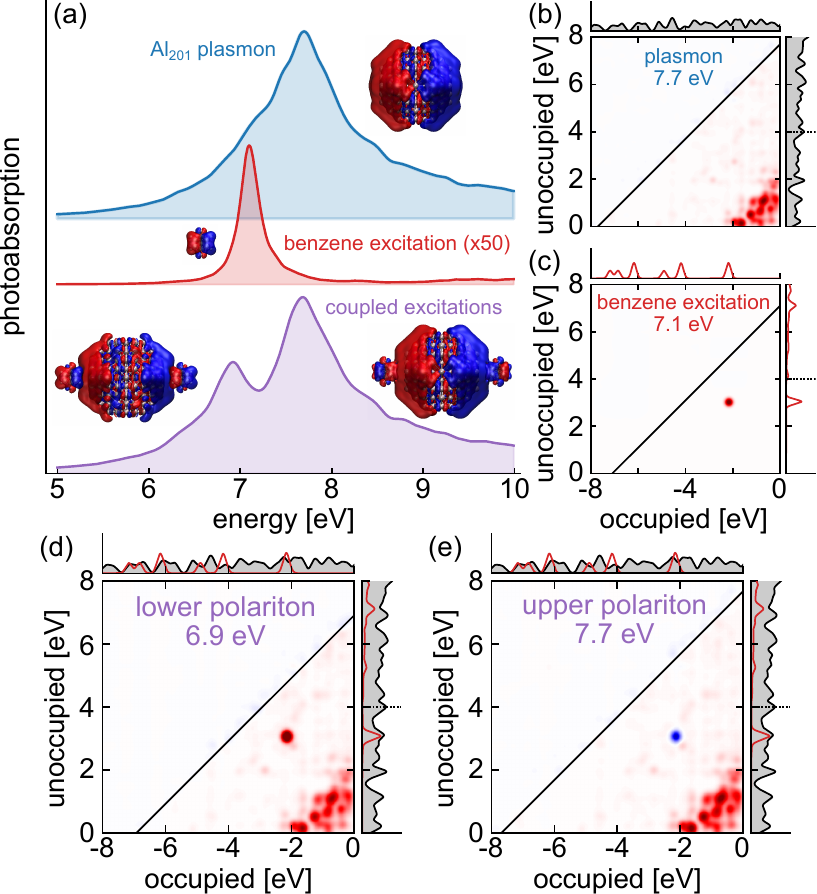}
\caption{Constituent elements of the strongly coupled system.
    (a) Strong interaction between a collective excitation (plasmon) of an \ce{Al201} nanoparticle and a molecular resonance of benzene results in the formation of lower and upper polariton states.
    Insets show the induced charge densities at resonances.
    The spectra are to scale but offset for clarity.
    (b-e) Transition contribution maps (TCMs) of the (b) plasmon and (c) molecular resonance show, respectively, the collective and discrete nature of these resonances.
    TCMs for (d) the lower polariton (\unit[6.92]{eV}) and (e) the upper polariton (\unit[7.68]{eV}) show the mix of plasmon and molecular states.
    The alignment of the molecular transition (ca.~$-2$ to \unit[3]{eV}) with respect to the plasmon in the LP/UP show clear symmetric and antisymmetric character.
    The induced energy densities at these energies visualize the in-phase alignment of the densities at the LP and out-of-phase for the UP.
    In the DOS red marks the contribution from benzene, multiplied by 5 for visibility,
    and the vacuum level at around \unit[4]{eV} is marked by a dashed line.}
\label{fig::1}
\end{figure}

We focus on proof-of-principle plasmonic systems based on idealized aluminum (Al) nanoparticles.
The free-electron-like electronic structure of Al greatly simplifies the analysis in contrast to noble metals with $d$-electron-screened plasmons.
In addition, Al has attracted recent interest as an alternative plasmonic material \cite{2014_ACSNano_8_834_knight}.

Our model systems consist of Al nanoparticles (regular truncated octahedra with 201, 586, and 1289 atoms) and benzene molecules (see Fig.~S4).
The \ce{Al201} nanoparticle exhibits a plasmon resonance at $\hbar\omega_{\mathrm{Al}}=\unit[7.7]{eV}$ (Fig.~\ref{fig::1}a), whose collective nature is recognizable in the TCM (Fig.~\ref{fig::1}b) \cite{2017_JCTC_13_4779_rossi}.
The benzene molecule has a doubly-degenerate resonance at $\hbar\omega_{\mathrm{B}}= \unit[7.1]{eV}$ with a transition dipole moment $\mu_{1}=\unit[4.45]{D}$ (Figs.~\ref{fig::1}a, S1).
When the nanoparticle and benzene molecules are placed in proximity (bottom part of Fig.~\ref{fig::1}a) their resonances couple and form two polaritons.
The lower (at \unit[6.92]{eV}) and upper (at \unit[7.68]{eV}) polaritons are, respectively, mixed symmetric and antisymmetric plasmon-molecule states as apparent in the TCMs of the coupled system (Fig.~\ref{fig::1}de).
A comparison with the TCMs of the uncoupled constituents (Fig.~\ref{fig::1}bc) reveals the individual contributions to the polaritons.
Transitions from just below the Fermi energy to just above correspond to the plasmon, while the contribution from about $-2$ to \unit[3]{eV} originates from the molecular exciton.
Crucially, the two polaritonic states differ with respect to the sign of the molecular contributions.
In case of the lower polariton (LP) plasmonic and molecular transitions are \emph{in-phase}, while for the upper polariton (UP) they are \emph{out-of-phase}.
This respectively symmetric and antisymmetric combination is clearly visible in the induced densities.
At the LP the dipoles of both particle and molecules are parallel, while they are antiparallel at the UP.
Such an arrangement is archetypal for a strongly coupled system.
This observation is valid also for all other calculated spectra, indicating the presence of strong or near-strong coupling already in the case of a single molecule placed \unit[3]{\AA} away from the \ce{Al201} particle.
We note here, that the density of states of the molecular elements in the coupled system does not significantly differ from their uncoupled states up to and including the LUMO (Fig.~S2).

\begin{figure}
\centering
\includegraphics{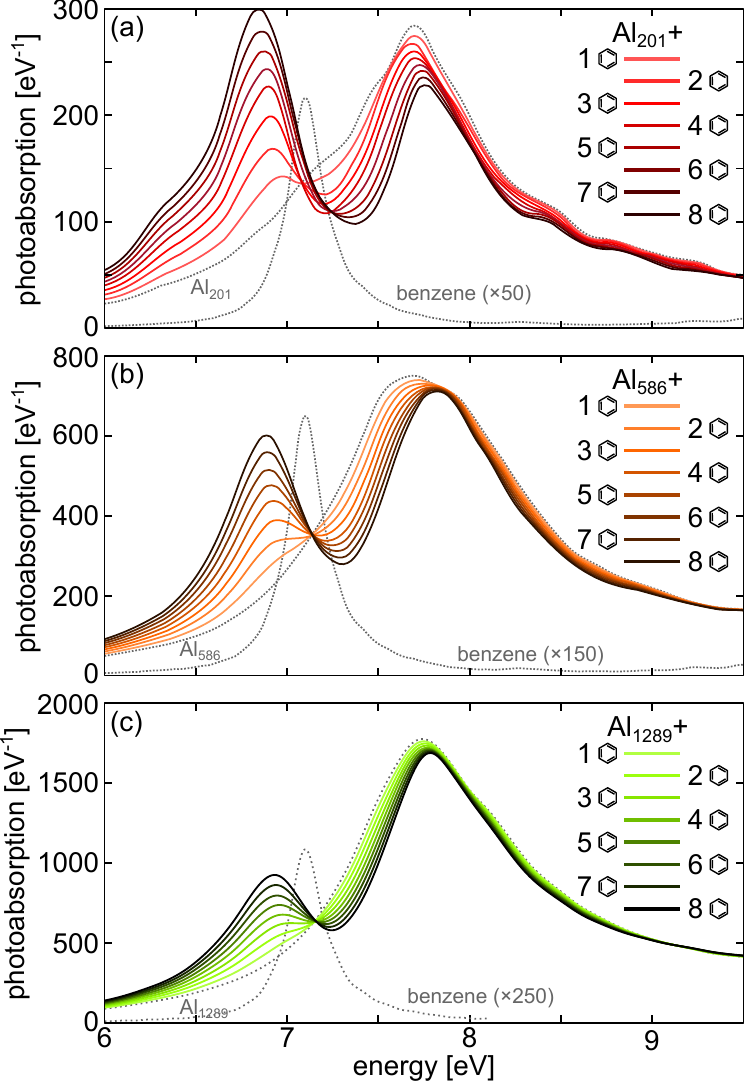}
\caption{
    Photoabsorption spectra for benzene molecules coupled to the plasmon of (a) \ce{Al201}, (b) \ce{Al586} and (c) \ce{Al1289} as a function of the number $N$ of benzene molecules (solid lines) showing clearly separated LP and UP, the splitting of which increases with $N$ (\unit[3]{\AA} separation).
    Spectra of the bare Al particles and benzene molecule are added for reference (dotted lines).}
\label{fig::2}
\end{figure}

To determine the fidelity of TDDFT for modeling strong coupling, we calculated the photoabsorption spectra of \ce{Al201}, \ce{Al586}, and \ce{Al1289} coupled to $N$ benzene molecules, whose dipole moment is aligned collinearly with that of the plasmon vacuum field.
The molecules are positioned at the corners of two opposing \{100\} facets of the particles (Fig.~S4), for simplicity neglecting relaxation.
The coupled systems exhibit the emergence of the LP and UP modes, the splitting of which increases with the number of molecules, see Fig.~\ref{fig::2} (TCMs show qualitatively identical symmetric/antisymmetric mixtures of plasmonic/molecular states, see Fig.~S3).
The Rabi splitting $2\Omega$ of the absorption spectra for \ce{Al201} and one benzene molecule equals \unit[730]{meV}, a value comparable to the plasmon width.
A non-negligible part of the UP/LP separation originates, however, from the plasmon-exciton detuning and the coupling strength $g$ is \unit[200]{meV} for one benzene, a value smaller than the geometrical mean of the plasmon and exciton widths (\unit[300]{meV}).
The strong coupling condition is, however, fulfilled for $N\ge3$ for \ce{Al201} and for still larger $N$ for larger particles.

\begin{figure}
\centering
\includegraphics{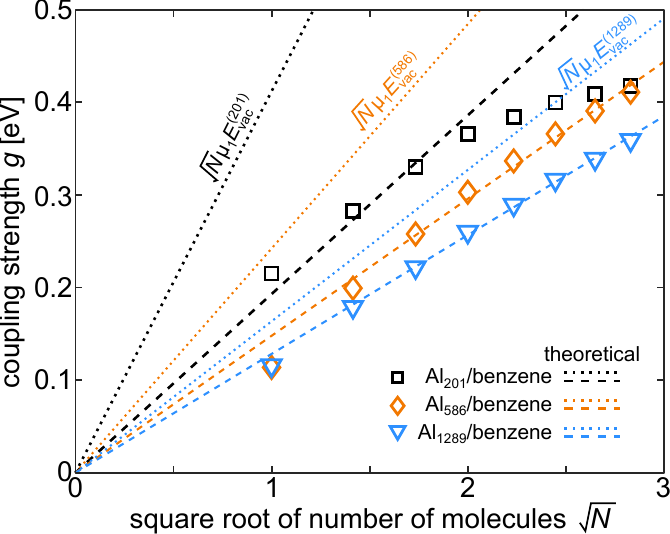}
\caption{
    Benzene-\ce{Al201}/\ce{Al586}/\ce{Al1289} coupling strength (a) as a function of the number of molecules placed \unit[3]{\AA} above the nanoparticle.
    The dotted lines mark the ideal theoretical coupling strengths $g=\sqrt{N}\mu_{1}E_{\mathrm{vac}}$.
    The symbols denote fitted $g$ values from the TDDFT data of Fig.~\ref{fig::2}, while the dashed lines mark the ideal dependence multiplied with the efficiency factor $\eta$ (see Fig.~\ref{fig::4}a and its in-text description for details) with $\eta_{201}= 0.47$, $\eta_{586}=0.61$, and $\eta_{1289}=0.78$.}
\label{fig::3}
\end{figure}

One of the signatures of strong coupling is the square-root dependence of $g$ on the number $N$ of identical excitons with transition dipole moment $\mu_{1}$ interacting with an optical cavity in the standard QED expression,
\begin{equation}
    g(N)=\sqrt{N}\mu_{1}E_{\mathrm{vac}},
    \label{eq:qed}
\end{equation}
where $E_{\mathrm{vac}}=\sqrt{\hbar\omega_{\mathrm{Al}}/2\varepsilon\varepsilon_{0}V}$ is the vacuum field and $V$ is the volume of the electromagnetic mode \cite{NatPhys_2_81_khitrova}.
Using the \ce{Al201} particle as an example, the expected scaling  is shown in Fig.~\ref{fig::3} by the black dotted line using $\mu_{1}$ of a benzene molecule and $E_{\mathrm{vac}}^{(201)}=\unitfrac[4.45]{V}{nm}$ with $\hbar\omega_{\mathrm{Al}}=\unit[7.7]{eV}$ and $V=\unit[3.3]{nm}^{3}$ corresponding to the volume of the \ce{Al201} particle, which is an adequate estimate for a nanometer-sized particle \cite{OL_35_4208_koenderink}.
To estimate the coupling strengths $g$ from our first-principles data, we fit the absorption spectra with a coupled harmonic oscillator model \cite{OE_18_23633_wu} (see Sec.~S4 in the SI and Figs.~S5-S8), yielding the $g$-values shown by squares in Fig.~\ref{fig::3}.
We notice that the coupling strength values obtained using the standard QED expression in Eq.~\eqref{eq:qed},
which is insensitive to spatial variations in the electric field, are larger than the ones derived from coupled oscillator fitting.
A similar observation holds for the larger particles, except that due to a larger volume vacuum field and coupling strengths are reduced.

\begin{figure}
\centering
\includegraphics[width=8.5cm]{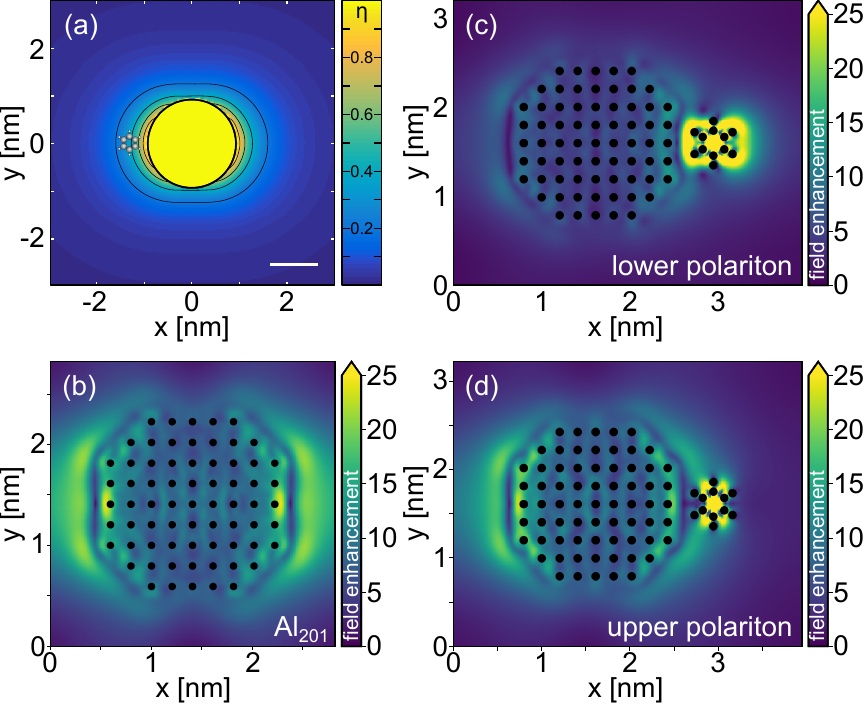}
\caption{
    (a) Spatial map of the coupling efficiency (normalized mode volume, calculated using classical electromagnetism) shows that maximum coupling occurs if benzene is inside the \ce{Al201} particle and decreases rapidly away from its surface.
    At the position of benzene the coupling efficiency is about 30 to 40\% of the maximum value. The contours are spaced every 0.2.
    (b-d) Field enhancement from TDDFT calculations at, respectively, the \ce{Al201} resonance and the \ce{Al201}-benzene LP and UP (with benzene located at the center of the facet).
    Benzene focuses the electric field, perturbing the nanoplasmonic cavity and increasing the coupling efficiency beyond the expectation based on the field at the bare resonance. \cite{2016_OpEx_24_20373_yang}
}
\label{fig::4}
\end{figure}

Equation~\eqref{eq:qed} is accurate under the condition that all excitonic modes are coherently coupled with the same, maximal rate to the cavity.
This is true for structures such as Fabry-Pérot cavities, photonic crystal slabs, or micropillars, in which the anti-node of the mode is accessible to excitonic modes of molecules or semiconductors by precise placement via trapping or doping \cite{NatPhys_2_81_khitrova, 2018_ASCP_5_24_baranov}.
For isolated plasmonic particles the maximum of its mode is, however, typically inside the particle \cite{OL_35_4208_koenderink}, such as in the present case.
Consequently, the molecular coherent dipole moment $\mu_{\mathrm{coh}}$ interacting with the cavity is smaller than its maximum value of \unit[4.45]{D}.
It is in the range 2 to \unit[4]{D} and increases with the number of Al atoms in the considered range of particle volumes.
The reduction of efficiency is typically expressed in terms of an efficiency factor $\eta$ determined by the ratio of the mode energy density at the position of the exciton to its maximum value \cite{2016_OpEx_24_20373_yang}.

Before extracting efficiency factors $\eta$ from the TDDFT data, we estimate their values on the basis of classical electromagnetic calculations, which are acceptable at a semi-quantitative level for picocavities \cite{2016_Sci_354_726_benz}, with a local Drude-permittivity tailored to match the TDDFT absorption spectrum for \ce{Al201}.
We find that in the spatial region occupied by the benzene molecule $\eta$ ranges from 0.2 to 0.4 (Fig.~\ref{fig::4}a), due to the rapid decrease of the plasmon-induced electric field.
However, such a low value of $\eta$ predicted by calculating the mode profile of a bare cavity does not hold in practice. It is known that the presence of a material with a refractive index greater than the background causes additional field localization, increasing the coupling strength \cite{2016_OpEx_24_20373_yang}.
While a single benzene molecule cannot be meaningfully described in terms of a refractive index, it has the identical effect.
This is confirmed by the electric field enhancements from TDDFT (Fig.~\ref{fig::4}e-g).
In comparison to the induced field of the bare plasmon, in the LP/UP of the coupled system benzene focuses the electric field, modifying the cavity and, consequently,  its vacuum field \cite{2016_OpEx_24_20373_yang}.

\begin{figure}
\centering
\includegraphics[width=8.5cm]{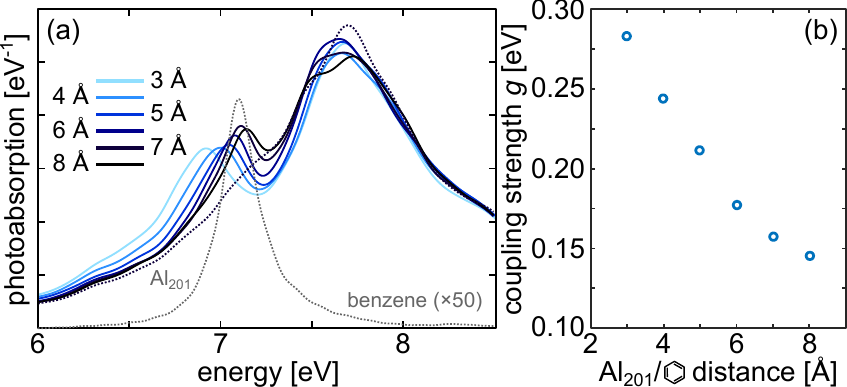}
\caption{
    (a) Photoabsorption spectra for two benzene molecules coupled to the \ce{Al201} particle, showing weaker splitting with increasing separation.
    (b) Coupling strength for two benzene molecules as a function of distance from the \ce{Al201} particle, showing the expected decrease due to the decreasing overlap integral between exciton and mode volume.}
\label{fig::5}
\end{figure}

We now turn to the extraction of the coupling efficiencies $\eta$ from the TDDFT-derived coupling strengths. The efficiencies modify the slope predicted by Eq.~\eqref{eq:qed} and the resulting dependencies are shown as dashed lines in Fig.~\ref{fig::3}.
$\eta_{201}= 0.47$ gives very good agreement for $N\le3$ (note the dashed black line marking the theoretical dependence $g=\eta_{201}\sqrt{N}\mu_{1}E_{\mathrm{vac}}$ in Fig.~\ref{fig::3}).
For $N > 3$ the calculated coupling strengths fall, however, below the slope set by $\eta_{201}$.
Such a deviation for large $N$ is not observed for the two larger particles, which, respectively, have efficiency factors of $\eta_{586} = 0.61$ and $\eta_{1289} = 0.78$.
These larger efficiencies imply that the plasmonic modes of the larger particles are used more efficiently than in the case of \ce{Al201}.
Minor deviations for the two larger nanoparticles are seen for $N<3$. However, they are probably caused by inaccurate fitting of the absorption spectra, as the LP/UP are not very clearly defined.

The efficiency factors obtained from classical electromagnetic and TDDFT simulations are consistent.
We note that in both scenarios the addition of molecules leads to a systematic red shift of the plasmon (Fig.~S9). This shift is the largest for \ce{Al201} and scales inversely with particle volume in agreement with predictions \cite{OE_20_524_tja}, further supporting the validity of our calculations.
This effect goes hand-in-hand with strong coupling and complicates the analysis.
The calculations yield a $\sqrt{N}$ dependence of the coupling strength for the larger particles (\ce{Al586}, \ce{Al1289}).
For the smallest particle (\ce{Al201}) the efficiency, however, varies with the number of benzene molecules for $N\gtrsim3$.
In this case, the volume occupied by the molecules is non-negligible relative to the particle volume and every additional molecule changes the properties of the nanoscale cavity and subsequently modifies the coupling.

We will now interpret these results from the perspective of classical electromagnetism.
The molecules can be represented as a polarizable background with excitons modelled by Lorentzians \cite{ACSPhoton_1_454_tja}.
This background can be treated as part of the cavity and consequently modifies its local optical density of states (LDoS) \cite{2016_OpEx_24_20373_yang, JPCC_119_11858_zhang}.
This modification manifests itself as focusing of the electric field around the molecules, clearly visible for \ce{Al201} coupled with one benzene molecule (Fig.~\ref{fig::4}cd), in comparison to the resonance of \ce{Al201} alone (Fig.~\ref{fig::4}b).
These LDoS modifications cause an increase of the mode volume, decreasing (or at least not enhancing) the coupling strength per molecule.
Such changes of the cavity induced by adding molecules are not visible for larger particles, as for \ce{Al586} and \ce{Al1289} the coupling strengths follow the $\sqrt{N}$ dependence.

A final point we discuss is the effect of a diminishing overlap between the vacuum field of the cavity and the molecular exciton. It is cleanly demonstrated via the distance dependence of the response, which we performed for the \ce{Al201} particle coupled to two benzene molecules (Fig.~\ref{fig::5}a).
The Rabi splitting of LP/UP decreases with increasing separation while the TCMs for LP/UP show qualitatively identical symmetric/antisymmetric mixtures of plasmonic/molecular states.
Furthermore, the reduction of $\eta$ \cite{2016_OpEx_24_20373_yang} leads to a decrease of the coupling strength (Fig.~\ref{fig::5}b).

In conclusion, we demonstrated the suitability of first-principles TDDFT for studying strong coupling between plasmons and molecular excitons.
This approach allows us to capture relevant interactions at the atomic level.
This is important for studying coupling of molecules to picocavities slightly larger than the molecules themselves \cite{2016_Sci_354_726_benz, BaumbergNature2016}) as well as ultrastrong coupling that leads to modification of the molecular ground state \cite{PRX_5_041022_galego, 2015_FaradDisc_178_281_george}.
Furthermore, we have shown the degree to which the simple cQED description holds for small systems.
For small single-particle cavities, such as the \ce{Al201} particle considered here, the presence of molecules modifies the cavity and, consequently, its mode volume.
The mode volume, in turn, affects the coupling strength, whose value does not increase as quickly as predicted by QED.
This slower increase of $g$ is noticeable for the \ce{Al201} particle, but for larger sizes any deviations from Eq.~\eqref{eq:qed} become negligible.
Despite the deviations between a simple cQED formula and TDDFT calculations, the order of magnitude agreement between them is quite remarkable.
In particular, it is rather surprising that such a simple formula at all holds at such small scales.
Based on the obtained coupling strengths, calculated mode volumes and field enhancements, reaching single/few-molecule ultrastrong coupling for single-particle cavities may prove challenging.

We acknowledge financial support from the Swedish Research Council (VR), the Knut and Alice Wallenberg Foundation, and the Polish National Science Center via the project 2017/25/B/ST3/00744.
Computational resources were kindly provided by the ICM-UW (Grant~\#G55-6) and by the Swedish National Infrastructure for Computing at PDC, Stockholm.
TJA also acknowledges support from the HPC-Europa3 program via project HPC17XPO18.

\section*{Methods}

\textbf{Computational details.}
The DFT and TDDFT calculations were carried out using the Perdew-Burke-Ernzerhof (PBE) \cite{1996_PRL_77_3865_perdew, *1997_PRL_78_1396_perdew} exchange-correlation functional in the adiabatic limit.
The spectra are calculated using the $\delta$-kick technique \cite{1996_PRB_54_4484_yabana} in the linear-response regime and employing the dipole approximation for light-matter interaction.
The default projector augmented-wave (PAW) \cite{1994_PRB_50_17953_blochl} data sets and double-$\zeta$ polarized (dzp) basis sets provided in GPAW were used.
The dzp basis set of Al includes diffuse 3p functions, which are important for describing plasmon resonances \cite{2015_JChemPhys_142_094114_rossi}.
To minimize spurious effects due to the basis-set superposition error, the so-called ghost-atom approach was used separately for each nanoparticle size to keep the total system basis set as intact as possible despite the changing number of surrounding molecules.
In general, while the used basis sets might not be adequate for yielding numerical values at the complete-basis-set limit, they are expected to be sufficient for the purposes of the present work.
A grid spacing parameter of \unit[0.3]{\AA} was chosen to represent densities and potentials, and the molecules/particles were surrounded by a vacuum region of at least \unit[6]{\AA}.
The Hartree potential was evaluated on a larger grid with at least \unit[120]{\AA} vacuum around the system and a coarser grid spacing of \unit[1.2]{\AA}, and subsequently refined to the original grid.
For the time propagation, we used a time step of $\Delta t = \unit[15]{as}$ and total propagation time of at least $T = \unit[30]{fs}$, which is sufficient for the used Lorentzian spectral broadening with $\eta = \unit[0.1]{eV}$ corresponding to a full width at half-maximum of \unit[0.2]{eV}.

\textbf{Transition contribution map (TCM).}
A TCM is used for visualizing the Kohn-Sham (KS) electron-hole transition contributions to photoabsorption.
Briefly, the photoabsorption cross-section $S(\omega)$ is expressed in the basis of occupied ($i$) and unoccupied ($a$) KS states as $S(\omega) = \sum_{ia}S_{ia}(\omega)$ \cite{2017_JCTC_13_4779_rossi}.
In TCM, the elements of the matrix $S_{ia}(\omega)$ at a chosen resonance energy are plotted on a Gaussian-broadened two-dimensional plane spanned by the energy axes for occupied ($\varepsilon_{o}$) and unoccupied ($\varepsilon_{u}$) KS states.
See Ref.~\onlinecite{2017_JCTC_13_4779_rossi} for a detailed description of TCM construction.
The maps are augmented with the corresponding densities of states, and a diagonal line is drawn to indicate the KS eigenvalue difference corresponding to $\omega$.

\addtolength{\textheight}{1em}

\bibliography{main}

\begin{thebibliography}{59}%
\makeatletter
\providecommand \@ifxundefined [1]{%
 \@ifx{#1\undefined}
}%
\providecommand \@ifnum [1]{%
 \ifnum #1\expandafter \@firstoftwo
 \else \expandafter \@secondoftwo
 \fi
}%
\providecommand \@ifx [1]{%
 \ifx #1\expandafter \@firstoftwo
 \else \expandafter \@secondoftwo
 \fi
}%
\providecommand \natexlab [1]{#1}%
\providecommand \enquote  [1]{``#1''}%
\providecommand \bibnamefont  [1]{#1}%
\providecommand \bibfnamefont [1]{#1}%
\providecommand \citenamefont [1]{#1}%
\providecommand \href@noop [0]{\@secondoftwo}%
\providecommand \href [0]{\begingroup \@sanitize@url \@href}%
\providecommand \@href[1]{\@@startlink{#1}\@@href}%
\providecommand \@@href[1]{\endgroup#1\@@endlink}%
\providecommand \@sanitize@url [0]{\catcode `\\12\catcode `\$12\catcode
  `\&12\catcode `\#12\catcode `\^12\catcode `\_12\catcode `\%12\relax}%
\providecommand \@@startlink[1]{}%
\providecommand \@@endlink[0]{}%
\providecommand \url  [0]{\begingroup\@sanitize@url \@url }%
\providecommand \@url [1]{\endgroup\@href {#1}{\urlprefix }}%
\providecommand \urlprefix  [0]{URL }%
\providecommand \Eprint [0]{\href }%
\providecommand \doibase [0]{http://dx.doi.org/}%
\providecommand \selectlanguage [0]{\@gobble}%
\providecommand \bibinfo  [0]{\@secondoftwo}%
\providecommand \bibfield  [0]{\@secondoftwo}%
\providecommand \translation [1]{[#1]}%
\providecommand \BibitemOpen [0]{}%
\providecommand \bibitemStop [0]{}%
\providecommand \bibitemNoStop [0]{.\EOS\space}%
\providecommand \EOS [0]{\spacefactor3000\relax}%
\providecommand \BibitemShut  [1]{\csname bibitem#1\endcsname}%
\let\auto@bib@innerbib\@empty
\bibitem [{\citenamefont {Scully}\ and\ \citenamefont
  {Zubairy}(1997)}]{scully_zubairy_1997}%
  \BibitemOpen
  \bibfield  {author} {\bibinfo {author} {\bibfnamefont {M.~O.}\ \bibnamefont
  {Scully}}\ and\ \bibinfo {author} {\bibfnamefont {M.~S.}\ \bibnamefont
  {Zubairy}},\ }\href {\doibase 10.1017/CBO9780511813993} {\emph {\bibinfo
  {title} {Quantum Optics}}}\ (\bibinfo  {publisher} {Cambridge University
  Press},\ \bibinfo {year} {1997})\BibitemShut {NoStop}%
\bibitem [{\citenamefont {T\"orm\"a}\ and\ \citenamefont
  {Barnes}(2015)}]{RepProgPhys_78_013901_torma}%
  \BibitemOpen
  \bibfield  {author} {\bibinfo {author} {\bibfnamefont {P.}~\bibnamefont
  {T\"orm\"a}}\ and\ \bibinfo {author} {\bibfnamefont {W.~L.}\ \bibnamefont
  {Barnes}},\ }\href {\doibase 10.1088/0034-4885/78/1/013901} {\bibfield
  {journal} {\bibinfo  {journal} {Rep. Prog. Phys.}\ }\textbf {\bibinfo
  {volume} {78}},\ \bibinfo {pages} {013901} (\bibinfo {year}
  {2015})}\BibitemShut {NoStop}%
\bibitem [{\citenamefont {Baranov}\ \emph {et~al.}(2018)\citenamefont
  {Baranov}, \citenamefont {Wers\"all}, \citenamefont {Cuadra}, \citenamefont
  {Antosiewicz},\ and\ \citenamefont {Shegai}}]{2018_ASCP_5_24_baranov}%
  \BibitemOpen
  \bibfield  {author} {\bibinfo {author} {\bibfnamefont {D.~G.}\ \bibnamefont
  {Baranov}}, \bibinfo {author} {\bibfnamefont {M.}~\bibnamefont {Wers\"all}},
  \bibinfo {author} {\bibfnamefont {J.}~\bibnamefont {Cuadra}}, \bibinfo
  {author} {\bibfnamefont {T.~J.}\ \bibnamefont {Antosiewicz}}, \ and\ \bibinfo
  {author} {\bibfnamefont {T.}~\bibnamefont {Shegai}},\ }\href {\doibase
  10.1021/acsphotonics.7b00674} {\bibfield  {journal} {\bibinfo  {journal} {ACS
  Photonics}\ }\textbf {\bibinfo {volume} {5}},\ \bibinfo {pages} {24}
  (\bibinfo {year} {2018})}\BibitemShut {NoStop}%
\bibitem [{\citenamefont {Kockum}\ \emph {et~al.}(2019)\citenamefont {Kockum},
  \citenamefont {Miranowicz}, \citenamefont {De~Liberato}, \citenamefont
  {Savasta},\ and\ \citenamefont {Nori}}]{NatRevPhys_1_19_frisk}%
  \BibitemOpen
  \bibfield  {author} {\bibinfo {author} {\bibfnamefont {A.~F.}\ \bibnamefont
  {Kockum}}, \bibinfo {author} {\bibfnamefont {A.}~\bibnamefont {Miranowicz}},
  \bibinfo {author} {\bibfnamefont {S.}~\bibnamefont {De~Liberato}}, \bibinfo
  {author} {\bibfnamefont {S.}~\bibnamefont {Savasta}}, \ and\ \bibinfo
  {author} {\bibfnamefont {F.}~\bibnamefont {Nori}},\ }\href {\doibase
  10.1038/s42254-018-0006-2} {\bibfield  {journal} {\bibinfo  {journal} {Nat.
  Rev. Phys.}\ }\textbf {\bibinfo {volume} {1}},\ \bibinfo {pages} {19}
  (\bibinfo {year} {2019})}\BibitemShut {NoStop}%
\bibitem [{\citenamefont {Yoshie}\ \emph {et~al.}(2004)\citenamefont {Yoshie},
  \citenamefont {Scherer}, \citenamefont {Hendrickson}, \citenamefont
  {Khitrova}, \citenamefont {Gibbs}, \citenamefont {Rupper}, \citenamefont
  {Ell}, \citenamefont {Shchekin},\ and\ \citenamefont
  {Deppe}}]{Nature_432_200_yoshie}%
  \BibitemOpen
  \bibfield  {author} {\bibinfo {author} {\bibfnamefont {T.}~\bibnamefont
  {Yoshie}}, \bibinfo {author} {\bibfnamefont {A.}~\bibnamefont {Scherer}},
  \bibinfo {author} {\bibfnamefont {J.}~\bibnamefont {Hendrickson}}, \bibinfo
  {author} {\bibfnamefont {G.}~\bibnamefont {Khitrova}}, \bibinfo {author}
  {\bibfnamefont {H.~M.}\ \bibnamefont {Gibbs}}, \bibinfo {author}
  {\bibfnamefont {G.}~\bibnamefont {Rupper}}, \bibinfo {author} {\bibfnamefont
  {C.}~\bibnamefont {Ell}}, \bibinfo {author} {\bibfnamefont {O.~B.}\
  \bibnamefont {Shchekin}}, \ and\ \bibinfo {author} {\bibfnamefont {D.~G.}\
  \bibnamefont {Deppe}},\ }\href {\doibase 10.1038/nature03119} {\bibfield
  {journal} {\bibinfo  {journal} {Nature}\ }\textbf {\bibinfo {volume} {432}},\
  \bibinfo {pages} {200} (\bibinfo {year} {2004})}\BibitemShut {NoStop}%
\bibitem [{\citenamefont {Birnbaum}\ \emph {et~al.}(2005)\citenamefont
  {Birnbaum}, \citenamefont {Boca}, \citenamefont {Miller}, \citenamefont
  {Boozer}, \citenamefont {Northup},\ and\ \citenamefont
  {Kimble}}]{Nature_436_87_birnbaum}%
  \BibitemOpen
  \bibfield  {author} {\bibinfo {author} {\bibfnamefont {K.~M.}\ \bibnamefont
  {Birnbaum}}, \bibinfo {author} {\bibfnamefont {A.}~\bibnamefont {Boca}},
  \bibinfo {author} {\bibfnamefont {R.}~\bibnamefont {Miller}}, \bibinfo
  {author} {\bibfnamefont {A.~D.}\ \bibnamefont {Boozer}}, \bibinfo {author}
  {\bibfnamefont {T.~E.}\ \bibnamefont {Northup}}, \ and\ \bibinfo {author}
  {\bibfnamefont {H.~J.}\ \bibnamefont {Kimble}},\ }\href {\doibase
  10.1038/nature03804} {\bibfield  {journal} {\bibinfo  {journal} {Nature}\
  }\textbf {\bibinfo {volume} {436}},\ \bibinfo {pages} {87} (\bibinfo {year}
  {2005})}\BibitemShut {NoStop}%
\bibitem [{\citenamefont {Kasprzak}\ \emph {et~al.}(2006)\citenamefont
  {Kasprzak}, \citenamefont {Richard}, \citenamefont {Kundermann},
  \citenamefont {Baas}, \citenamefont {Jeambrun}, \citenamefont {Keeling},
  \citenamefont {Marchetti}, \citenamefont {Szymanska}, \citenamefont {Andre},
  \citenamefont {Staehli}, \citenamefont {Savona}, \citenamefont {Littlewood},
  \citenamefont {Deveaud},\ and\ \citenamefont
  {Dang}}]{Nature_443_409_kasprzak}%
  \BibitemOpen
  \bibfield  {author} {\bibinfo {author} {\bibfnamefont {J.}~\bibnamefont
  {Kasprzak}}, \bibinfo {author} {\bibfnamefont {M.}~\bibnamefont {Richard}},
  \bibinfo {author} {\bibfnamefont {S.}~\bibnamefont {Kundermann}}, \bibinfo
  {author} {\bibfnamefont {A.}~\bibnamefont {Baas}}, \bibinfo {author}
  {\bibfnamefont {P.}~\bibnamefont {Jeambrun}}, \bibinfo {author}
  {\bibfnamefont {J.~M.~J.}\ \bibnamefont {Keeling}}, \bibinfo {author}
  {\bibfnamefont {F.~M.}\ \bibnamefont {Marchetti}}, \bibinfo {author}
  {\bibfnamefont {M.~H.}\ \bibnamefont {Szymanska}}, \bibinfo {author}
  {\bibfnamefont {R.}~\bibnamefont {Andre}}, \bibinfo {author} {\bibfnamefont
  {J.~L.}\ \bibnamefont {Staehli}}, \bibinfo {author} {\bibfnamefont
  {V.}~\bibnamefont {Savona}}, \bibinfo {author} {\bibfnamefont {P.~B.}\
  \bibnamefont {Littlewood}}, \bibinfo {author} {\bibfnamefont
  {B.}~\bibnamefont {Deveaud}}, \ and\ \bibinfo {author} {\bibfnamefont
  {L.~S.}\ \bibnamefont {Dang}},\ }\href {\doibase 10.1038/nature05131}
  {\bibfield  {journal} {\bibinfo  {journal} {Nature}\ }\textbf {\bibinfo
  {volume} {443}},\ \bibinfo {pages} {409} (\bibinfo {year}
  {2006})}\BibitemShut {NoStop}%
\bibitem [{\citenamefont {Hutchison}\ \emph {et~al.}(2012)\citenamefont
  {Hutchison}, \citenamefont {Schwartz}, \citenamefont {Genet}, \citenamefont
  {Devaux},\ and\ \citenamefont {Ebbesen}}]{AngChem_51_1592_ebbesen}%
  \BibitemOpen
  \bibfield  {author} {\bibinfo {author} {\bibfnamefont {J.~A.}\ \bibnamefont
  {Hutchison}}, \bibinfo {author} {\bibfnamefont {T.}~\bibnamefont {Schwartz}},
  \bibinfo {author} {\bibfnamefont {C.}~\bibnamefont {Genet}}, \bibinfo
  {author} {\bibfnamefont {E.}~\bibnamefont {Devaux}}, \ and\ \bibinfo {author}
  {\bibfnamefont {T.~W.}\ \bibnamefont {Ebbesen}},\ }\href {\doibase
  10.1002/anie.201107033} {\bibfield  {journal} {\bibinfo  {journal} {Angew.
  Chem. Int. Ed.}\ }\textbf {\bibinfo {volume} {51}},\ \bibinfo {pages} {1592}
  (\bibinfo {year} {2012})}\BibitemShut {NoStop}%
\bibitem [{\citenamefont {Herrera}\ and\ \citenamefont
  {Spano}(2016)}]{PRL_116_238301_herrera}%
  \BibitemOpen
  \bibfield  {author} {\bibinfo {author} {\bibfnamefont {F.}~\bibnamefont
  {Herrera}}\ and\ \bibinfo {author} {\bibfnamefont {F.~C.}\ \bibnamefont
  {Spano}},\ }\href {\doibase 10.1103/PhysRevLett.116.238301} {\bibfield
  {journal} {\bibinfo  {journal} {Phys. Rev. Lett.}\ }\textbf {\bibinfo
  {volume} {116}},\ \bibinfo {pages} {238301} (\bibinfo {year}
  {2016})}\BibitemShut {NoStop}%
\bibitem [{\citenamefont {Galego}\ \emph {et~al.}(2016)\citenamefont {Galego},
  \citenamefont {Garcia-Vidal},\ and\ \citenamefont
  {Feist}}]{NComm_7_13841_galego}%
  \BibitemOpen
  \bibfield  {author} {\bibinfo {author} {\bibfnamefont {J.}~\bibnamefont
  {Galego}}, \bibinfo {author} {\bibfnamefont {F.~J.}\ \bibnamefont
  {Garcia-Vidal}}, \ and\ \bibinfo {author} {\bibfnamefont {J.}~\bibnamefont
  {Feist}},\ }\href {\doibase 10.1038/ncomms13841} {\bibfield  {journal}
  {\bibinfo  {journal} {Nat. Commun.}\ }\textbf {\bibinfo {volume} {7}},\
  \bibinfo {pages} {13841} (\bibinfo {year} {2016})}\BibitemShut {NoStop}%
\bibitem [{\citenamefont {Mart\'{i}nez-Mart\'{i}nez}\ \emph
  {et~al.}(2018)\citenamefont {Mart\'{i}nez-Mart\'{i}nez}, \citenamefont
  {Ribeiro}, \citenamefont {Campos-Gonz\'{a}lez-Angulo},\ and\ \citenamefont
  {Yuen-Zhou}}]{ACSP_5_167_martinez}%
  \BibitemOpen
  \bibfield  {author} {\bibinfo {author} {\bibfnamefont {L.~A.}\ \bibnamefont
  {Mart\'{i}nez-Mart\'{i}nez}}, \bibinfo {author} {\bibfnamefont {R.~F.}\
  \bibnamefont {Ribeiro}}, \bibinfo {author} {\bibfnamefont {J.}~\bibnamefont
  {Campos-Gonz\'{a}lez-Angulo}}, \ and\ \bibinfo {author} {\bibfnamefont
  {J.}~\bibnamefont {Yuen-Zhou}},\ }\href {\doibase
  10.1021/acsphotonics.7b00610} {\bibfield  {journal} {\bibinfo  {journal} {ACS
  Photonics}\ }\textbf {\bibinfo {volume} {5}},\ \bibinfo {pages} {167}
  (\bibinfo {year} {2018})}\BibitemShut {NoStop}%
\bibitem [{\citenamefont {Munkhbat}\ \emph {et~al.}(2018)\citenamefont
  {Munkhbat}, \citenamefont {Wers{\"a}ll}, \citenamefont {Baranov},
  \citenamefont {Antosiewicz},\ and\ \citenamefont
  {Shegai}}]{SciAdv_4_eaas9552_munkhbat}%
  \BibitemOpen
  \bibfield  {author} {\bibinfo {author} {\bibfnamefont {B.}~\bibnamefont
  {Munkhbat}}, \bibinfo {author} {\bibfnamefont {M.}~\bibnamefont
  {Wers{\"a}ll}}, \bibinfo {author} {\bibfnamefont {D.~G.}\ \bibnamefont
  {Baranov}}, \bibinfo {author} {\bibfnamefont {T.~J.}\ \bibnamefont
  {Antosiewicz}}, \ and\ \bibinfo {author} {\bibfnamefont {T.}~\bibnamefont
  {Shegai}},\ }\href {\doibase 10.1126/sciadv.aas9552} {\bibfield  {journal}
  {\bibinfo  {journal} {Sci. Adv.}\ }\textbf {\bibinfo {volume} {4}},\ \bibinfo
  {pages} {eaas9552} (\bibinfo {year} {2018})}\BibitemShut {NoStop}%
\bibitem [{\citenamefont {Feist}\ and\ \citenamefont
  {Garcia-Vidal}(2015)}]{PRL_114_196402_feist}%
  \BibitemOpen
  \bibfield  {author} {\bibinfo {author} {\bibfnamefont {J.}~\bibnamefont
  {Feist}}\ and\ \bibinfo {author} {\bibfnamefont {F.~J.}\ \bibnamefont
  {Garcia-Vidal}},\ }\href {\doibase 10.1103/PhysRevLett.114.196402} {\bibfield
   {journal} {\bibinfo  {journal} {Phys. Rev. Lett.}\ }\textbf {\bibinfo
  {volume} {114}},\ \bibinfo {pages} {196402} (\bibinfo {year}
  {2015})}\BibitemShut {NoStop}%
\bibitem [{\citenamefont {Zhong}\ \emph {et~al.}(2016)\citenamefont {Zhong},
  \citenamefont {Chervy}, \citenamefont {Wang}, \citenamefont {George},
  \citenamefont {Thomas}, \citenamefont {Hutchison}, \citenamefont {Devaux},
  \citenamefont {Genet},\ and\ \citenamefont
  {Ebbesen}}]{AngChem_55_6202_zhong}%
  \BibitemOpen
  \bibfield  {author} {\bibinfo {author} {\bibfnamefont {X.}~\bibnamefont
  {Zhong}}, \bibinfo {author} {\bibfnamefont {T.}~\bibnamefont {Chervy}},
  \bibinfo {author} {\bibfnamefont {S.}~\bibnamefont {Wang}}, \bibinfo {author}
  {\bibfnamefont {J.}~\bibnamefont {George}}, \bibinfo {author} {\bibfnamefont
  {A.}~\bibnamefont {Thomas}}, \bibinfo {author} {\bibfnamefont {J.~A.}\
  \bibnamefont {Hutchison}}, \bibinfo {author} {\bibfnamefont {E.}~\bibnamefont
  {Devaux}}, \bibinfo {author} {\bibfnamefont {C.}~\bibnamefont {Genet}}, \
  and\ \bibinfo {author} {\bibfnamefont {T.~W.}\ \bibnamefont {Ebbesen}},\
  }\href {\doibase 10.1002/anie.201600428} {\bibfield  {journal} {\bibinfo
  {journal} {Angew. Chem. Int. Ed.}\ }\textbf {\bibinfo {volume} {55}},\
  \bibinfo {pages} {6202} (\bibinfo {year} {2016})}\BibitemShut {NoStop}%
\bibitem [{\citenamefont {Khitrova}\ \emph {et~al.}(2006)\citenamefont
  {Khitrova}, \citenamefont {Gibbs}, \citenamefont {Kira}, \citenamefont
  {Koch},\ and\ \citenamefont {Scherer}}]{NatPhys_2_81_khitrova}%
  \BibitemOpen
  \bibfield  {author} {\bibinfo {author} {\bibfnamefont {G.}~\bibnamefont
  {Khitrova}}, \bibinfo {author} {\bibfnamefont {M.}~\bibnamefont {Gibbs}},
  \bibinfo {author} {\bibfnamefont {M.}~\bibnamefont {Kira}}, \bibinfo {author}
  {\bibfnamefont {S.~W.}\ \bibnamefont {Koch}}, \ and\ \bibinfo {author}
  {\bibfnamefont {A.}~\bibnamefont {Scherer}},\ }\href {\doibase
  10.1038/nphys227} {\bibfield  {journal} {\bibinfo  {journal} {Nat. Phys.}\
  }\textbf {\bibinfo {volume} {2}},\ \bibinfo {pages} {81} (\bibinfo {year}
  {2006})}\BibitemShut {NoStop}%
\bibitem [{\citenamefont {Jaynes}\ and\ \citenamefont
  {Cummings}(1963)}]{ProcIEEE_51_89_haynes}%
  \BibitemOpen
  \bibfield  {author} {\bibinfo {author} {\bibfnamefont {E.~T.}\ \bibnamefont
  {Jaynes}}\ and\ \bibinfo {author} {\bibfnamefont {F.~W.}\ \bibnamefont
  {Cummings}},\ }\href {\doibase 10.1109/PROC.1963.1664} {\bibfield  {journal}
  {\bibinfo  {journal} {Proc. IEEE}\ }\textbf {\bibinfo {volume} {51}},\
  \bibinfo {pages} {89} (\bibinfo {year} {1963})}\BibitemShut {NoStop}%
\bibitem [{\citenamefont {Garraway}(2011)}]{PhilTransA_369_1137_garraway}%
  \BibitemOpen
  \bibfield  {author} {\bibinfo {author} {\bibfnamefont {B.~M.}\ \bibnamefont
  {Garraway}},\ }\href {\doibase 10.1098/rsta.2010.0333} {\bibfield  {journal}
  {\bibinfo  {journal} {Philos. Trans. Royal Soc. A}\ }\textbf {\bibinfo
  {volume} {369}},\ \bibinfo {pages} {1137} (\bibinfo {year}
  {2011})}\BibitemShut {NoStop}%
\bibitem [{\citenamefont {Galego}\ \emph {et~al.}(2015)\citenamefont {Galego},
  \citenamefont {Garcia-Vidal},\ and\ \citenamefont
  {Feist}}]{PRX_5_041022_galego}%
  \BibitemOpen
  \bibfield  {author} {\bibinfo {author} {\bibfnamefont {J.}~\bibnamefont
  {Galego}}, \bibinfo {author} {\bibfnamefont {F.~J.}\ \bibnamefont
  {Garcia-Vidal}}, \ and\ \bibinfo {author} {\bibfnamefont {J.}~\bibnamefont
  {Feist}},\ }\href {\doibase 10.1103/PhysRevX.5.041022} {\bibfield  {journal}
  {\bibinfo  {journal} {Phys. Rev. X}\ }\textbf {\bibinfo {volume} {5}},\
  \bibinfo {pages} {041022} (\bibinfo {year} {2015})}\BibitemShut {NoStop}%
\bibitem [{\citenamefont {Ruggenthaler}\ \emph {et~al.}(2014)\citenamefont
  {Ruggenthaler}, \citenamefont {Flick}, \citenamefont {Pellegrini},
  \citenamefont {Appel}, \citenamefont {Tokatly},\ and\ \citenamefont
  {Rubio}}]{PRA_90_012508_ruggenthaler}%
  \BibitemOpen
  \bibfield  {author} {\bibinfo {author} {\bibfnamefont {M.}~\bibnamefont
  {Ruggenthaler}}, \bibinfo {author} {\bibfnamefont {J.}~\bibnamefont {Flick}},
  \bibinfo {author} {\bibfnamefont {C.}~\bibnamefont {Pellegrini}}, \bibinfo
  {author} {\bibfnamefont {H.}~\bibnamefont {Appel}}, \bibinfo {author}
  {\bibfnamefont {I.~V.}\ \bibnamefont {Tokatly}}, \ and\ \bibinfo {author}
  {\bibfnamefont {A.}~\bibnamefont {Rubio}},\ }\href {\doibase
  10.1103/PhysRevA.90.012508} {\bibfield  {journal} {\bibinfo  {journal} {Phys.
  Rev. A}\ }\textbf {\bibinfo {volume} {90}},\ \bibinfo {pages} {012508}
  (\bibinfo {year} {2014})}\BibitemShut {NoStop}%
\bibitem [{\citenamefont {Ling~Luk}\ \emph {et~al.}(2017)\citenamefont
  {Ling~Luk}, \citenamefont {Feist}, \citenamefont {Toppari},\ and\
  \citenamefont {Groenhof}}]{JCTC_13_4324_ling}%
  \BibitemOpen
  \bibfield  {author} {\bibinfo {author} {\bibfnamefont {H.}~\bibnamefont
  {Ling~Luk}}, \bibinfo {author} {\bibfnamefont {J.}~\bibnamefont {Feist}},
  \bibinfo {author} {\bibfnamefont {J.~J.}\ \bibnamefont {Toppari}}, \ and\
  \bibinfo {author} {\bibfnamefont {G.}~\bibnamefont {Groenhof}},\ }\href
  {\doibase 10.1021/acs.jctc.7b00388} {\bibfield  {journal} {\bibinfo
  {journal} {J. Chem. Theory Comput.}\ }\textbf {\bibinfo {volume} {13}},\
  \bibinfo {pages} {4324} (\bibinfo {year} {2017})}\BibitemShut {NoStop}%
\bibitem [{\citenamefont {Neuman}\ \emph {et~al.}(2018)\citenamefont {Neuman},
  \citenamefont {Esteban}, \citenamefont {Casanova}, \citenamefont
  {Garcia-Vidal},\ and\ \citenamefont {Aizpurua}}]{NL_18_2358_neuman}%
  \BibitemOpen
  \bibfield  {author} {\bibinfo {author} {\bibfnamefont {T.}~\bibnamefont
  {Neuman}}, \bibinfo {author} {\bibfnamefont {R.}~\bibnamefont {Esteban}},
  \bibinfo {author} {\bibfnamefont {D.}~\bibnamefont {Casanova}}, \bibinfo
  {author} {\bibfnamefont {F.~J.}\ \bibnamefont {Garcia-Vidal}}, \ and\
  \bibinfo {author} {\bibfnamefont {J.}~\bibnamefont {Aizpurua}},\ }\href
  {\doibase 10.1021/acs.nanolett.7b05297} {\bibfield  {journal} {\bibinfo
  {journal} {Nano Lett.}\ }\textbf {\bibinfo {volume} {18}},\ \bibinfo {pages}
  {2358} (\bibinfo {year} {2018})}\BibitemShut {NoStop}%
\bibitem [{\citenamefont {Runge}\ and\ \citenamefont
  {Gross}(1984)}]{1984_PRL_52_997_runge}%
  \BibitemOpen
  \bibfield  {author} {\bibinfo {author} {\bibfnamefont {E.}~\bibnamefont
  {Runge}}\ and\ \bibinfo {author} {\bibfnamefont {E.~K.~U.}\ \bibnamefont
  {Gross}},\ }\href {\doibase 10.1103/PhysRevLett.52.997} {\bibfield  {journal}
  {\bibinfo  {journal} {Phys. Rev. Lett.}\ }\textbf {\bibinfo {volume} {52}},\
  \bibinfo {pages} {997} (\bibinfo {year} {1984})}\BibitemShut {NoStop}%
\bibitem [{\citenamefont {Hohenberg}\ and\ \citenamefont
  {Kohn}(1964)}]{1964_PhysRev_136_B864_hohenberg}%
  \BibitemOpen
  \bibfield  {author} {\bibinfo {author} {\bibfnamefont {P.}~\bibnamefont
  {Hohenberg}}\ and\ \bibinfo {author} {\bibfnamefont {W.}~\bibnamefont
  {Kohn}},\ }\href {\doibase 10.1103/PhysRev.136.B864} {\bibfield  {journal}
  {\bibinfo  {journal} {Phys. Rev.}\ }\textbf {\bibinfo {volume} {136}},\
  \bibinfo {pages} {B864} (\bibinfo {year} {1964})}\BibitemShut {NoStop}%
\bibitem [{\citenamefont {Kohn}\ and\ \citenamefont
  {Sham}(1965)}]{1965_PhysRev_140_A1133_kohn}%
  \BibitemOpen
  \bibfield  {author} {\bibinfo {author} {\bibfnamefont {W.}~\bibnamefont
  {Kohn}}\ and\ \bibinfo {author} {\bibfnamefont {L.~J.}\ \bibnamefont
  {Sham}},\ }\href {\doibase 10.1103/PhysRev.140.A1133} {\bibfield  {journal}
  {\bibinfo  {journal} {Phys. Rev.}\ }\textbf {\bibinfo {volume} {140}},\
  \bibinfo {pages} {A1133} (\bibinfo {year} {1965})}\BibitemShut {NoStop}%
\bibitem [{\citenamefont {Arenas}\ \emph {et~al.}(1996)\citenamefont {Arenas},
  \citenamefont {Woolley}, \citenamefont {Otero},\ and\ \citenamefont
  {Marcos}}]{1996_JPhysChem_100_3199_arenas}%
  \BibitemOpen
  \bibfield  {author} {\bibinfo {author} {\bibfnamefont {J.~F.}\ \bibnamefont
  {Arenas}}, \bibinfo {author} {\bibfnamefont {M.~S.}\ \bibnamefont {Woolley}},
  \bibinfo {author} {\bibfnamefont {J.~C.}\ \bibnamefont {Otero}}, \ and\
  \bibinfo {author} {\bibfnamefont {J.~I.}\ \bibnamefont {Marcos}},\ }\href
  {\doibase 10.1021/jp952240k} {\bibfield  {journal} {\bibinfo  {journal} {J.
  Phys. Chem.}\ }\textbf {\bibinfo {volume} {100}},\ \bibinfo {pages} {3199}
  (\bibinfo {year} {1996})}\BibitemShut {NoStop}%
\bibitem [{\citenamefont {Arenas}\ \emph {et~al.}(2002)\citenamefont {Arenas},
  \citenamefont {Soto}, \citenamefont {Tocon}, \citenamefont {Fernandez},
  \citenamefont {Otero},\ and\ \citenamefont
  {Marcos}}]{2002_JChemPhys_116_7207_arenas}%
  \BibitemOpen
  \bibfield  {author} {\bibinfo {author} {\bibfnamefont {J.~F.}\ \bibnamefont
  {Arenas}}, \bibinfo {author} {\bibfnamefont {J.}~\bibnamefont {Soto}},
  \bibinfo {author} {\bibfnamefont {I.~L.}\ \bibnamefont {Tocon}}, \bibinfo
  {author} {\bibfnamefont {D.~J.}\ \bibnamefont {Fernandez}}, \bibinfo {author}
  {\bibfnamefont {J.~C.}\ \bibnamefont {Otero}}, \ and\ \bibinfo {author}
  {\bibfnamefont {J.~I.}\ \bibnamefont {Marcos}},\ }\href {\doibase
  10.1063/1.1450542} {\bibfield  {journal} {\bibinfo  {journal} {J. Chem.
  Phys.}\ }\textbf {\bibinfo {volume} {116}},\ \bibinfo {pages} {7207}
  (\bibinfo {year} {2002})}\BibitemShut {NoStop}%
\bibitem [{\citenamefont {Bjerneld}\ \emph {et~al.}(2004)\citenamefont
  {Bjerneld}, \citenamefont {Svedberg}, \citenamefont {Johansson},\ and\
  \citenamefont {K\"all}}]{2004_JPhysChemA_108_4187_bjerneld}%
  \BibitemOpen
  \bibfield  {author} {\bibinfo {author} {\bibfnamefont {E.~J.}\ \bibnamefont
  {Bjerneld}}, \bibinfo {author} {\bibfnamefont {F.}~\bibnamefont {Svedberg}},
  \bibinfo {author} {\bibfnamefont {P.}~\bibnamefont {Johansson}}, \ and\
  \bibinfo {author} {\bibfnamefont {M.}~\bibnamefont {K\"all}},\ }\href
  {\doibase 10.1021/jp037004l} {\bibfield  {journal} {\bibinfo  {journal} {J.
  Phys. Chem. A}\ }\textbf {\bibinfo {volume} {108}},\ \bibinfo {pages} {4187}
  (\bibinfo {year} {2004})}\BibitemShut {NoStop}%
\bibitem [{\citenamefont {Zhao}\ \emph {et~al.}(2006)\citenamefont {Zhao},
  \citenamefont {Jensen},\ and\ \citenamefont
  {Schatz}}]{2006_JACS_128_2911_zhao}%
  \BibitemOpen
  \bibfield  {author} {\bibinfo {author} {\bibfnamefont {L.~L.}\ \bibnamefont
  {Zhao}}, \bibinfo {author} {\bibfnamefont {L.}~\bibnamefont {Jensen}}, \ and\
  \bibinfo {author} {\bibfnamefont {G.~C.}\ \bibnamefont {Schatz}},\ }\href
  {\doibase 10.1021/ja0556326} {\bibfield  {journal} {\bibinfo  {journal} {J.
  Am. Chem. Soc.}\ }\textbf {\bibinfo {volume} {128}},\ \bibinfo {pages} {2911}
  (\bibinfo {year} {2006})}\BibitemShut {NoStop}%
\bibitem [{\citenamefont {Jensen}\ \emph {et~al.}(2008)\citenamefont {Jensen},
  \citenamefont {Aikens},\ and\ \citenamefont
  {Schatz}}]{2008_ChemSocRev_37_1061_jensen}%
  \BibitemOpen
  \bibfield  {author} {\bibinfo {author} {\bibfnamefont {L.}~\bibnamefont
  {Jensen}}, \bibinfo {author} {\bibfnamefont {C.~M.}\ \bibnamefont {Aikens}},
  \ and\ \bibinfo {author} {\bibfnamefont {G.~C.}\ \bibnamefont {Schatz}},\
  }\href {\doibase 10.1039/B706023H} {\bibfield  {journal} {\bibinfo  {journal}
  {Chem. Soc. Rev.}\ }\textbf {\bibinfo {volume} {37}},\ \bibinfo {pages}
  {1061} (\bibinfo {year} {2008})}\BibitemShut {NoStop}%
\bibitem [{\citenamefont {Morton}\ and\ \citenamefont
  {Jensen}(2011)}]{2011_JChemPhys_135_1344103_morton}%
  \BibitemOpen
  \bibfield  {author} {\bibinfo {author} {\bibfnamefont {S.~M.}\ \bibnamefont
  {Morton}}\ and\ \bibinfo {author} {\bibfnamefont {L.}~\bibnamefont
  {Jensen}},\ }\href {\doibase 10.1063/1.3643381} {\bibfield  {journal}
  {\bibinfo  {journal} {J. Chem. Phys.}\ }\textbf {\bibinfo {volume} {135}},\
  \bibinfo {pages} {134103} (\bibinfo {year} {2011})}\BibitemShut {NoStop}%
\bibitem [{\citenamefont {Corni}\ and\ \citenamefont
  {Tomasi}(2001)}]{2001_ChemPhysLett_342_135_corni}%
  \BibitemOpen
  \bibfield  {author} {\bibinfo {author} {\bibfnamefont {S.}~\bibnamefont
  {Corni}}\ and\ \bibinfo {author} {\bibfnamefont {J.}~\bibnamefont {Tomasi}},\
  }\href {\doibase 10.1016/S0009-2614(01)00582-6} {\bibfield  {journal}
  {\bibinfo  {journal} {Chem. Phys. Lett.}\ }\textbf {\bibinfo {volume}
  {342}},\ \bibinfo {pages} {135} (\bibinfo {year} {2001})}\BibitemShut
  {NoStop}%
\bibitem [{\citenamefont {Lombardi}\ and\ \citenamefont
  {Birke}(2008)}]{2008_JPCC_112_5605_lombardi}%
  \BibitemOpen
  \bibfield  {author} {\bibinfo {author} {\bibfnamefont {J.}~\bibnamefont
  {Lombardi}}\ and\ \bibinfo {author} {\bibfnamefont {R.}~\bibnamefont
  {Birke}},\ }\href {\doibase 10.1021/jp800167v} {\bibfield  {journal}
  {\bibinfo  {journal} {J. Phys. Chem. C}\ }\textbf {\bibinfo {volume} {112}},\
  \bibinfo {pages} {5605} (\bibinfo {year} {2008})}\BibitemShut {NoStop}%
\bibitem [{\citenamefont {Shegai}\ \emph {et~al.}(2009)\citenamefont {Shegai},
  \citenamefont {Vaskevich}, \citenamefont {Rubinstein},\ and\ \citenamefont
  {Haran}}]{2009_JACS_131_14390_shegai}%
  \BibitemOpen
  \bibfield  {author} {\bibinfo {author} {\bibfnamefont {T.}~\bibnamefont
  {Shegai}}, \bibinfo {author} {\bibfnamefont {A.}~\bibnamefont {Vaskevich}},
  \bibinfo {author} {\bibfnamefont {I.}~\bibnamefont {Rubinstein}}, \ and\
  \bibinfo {author} {\bibfnamefont {G.}~\bibnamefont {Haran}},\ }\href
  {\doibase 10.1021/ja904480r} {\bibfield  {journal} {\bibinfo  {journal} {J.
  Am. Chem. Soc.}\ }\textbf {\bibinfo {volume} {131}},\ \bibinfo {pages}
  {14390} (\bibinfo {year} {2009})}\BibitemShut {NoStop}%
\bibitem [{\citenamefont {Flick}\ \emph {et~al.}(2017)\citenamefont {Flick},
  \citenamefont {Ruggenthaler}, \citenamefont {Appel},\ and\ \citenamefont
  {Rubio}}]{2017_PNAS_114_3026_flick}%
  \BibitemOpen
  \bibfield  {author} {\bibinfo {author} {\bibfnamefont {J.}~\bibnamefont
  {Flick}}, \bibinfo {author} {\bibfnamefont {M.}~\bibnamefont {Ruggenthaler}},
  \bibinfo {author} {\bibfnamefont {H.}~\bibnamefont {Appel}}, \ and\ \bibinfo
  {author} {\bibfnamefont {A.}~\bibnamefont {Rubio}},\ }\href {\doibase
  10.1073/pnas.1615509114} {\bibfield  {journal} {\bibinfo  {journal} {Proc.
  Nat. Acad. Sci. U.S.A.}\ }\textbf {\bibinfo {volume} {114}},\ \bibinfo
  {pages} {3026} (\bibinfo {year} {2017})}\BibitemShut {NoStop}%
\bibitem [{\citenamefont {Flick}\ \emph {et~al.}(2018)\citenamefont {Flick},
  \citenamefont {Rivera},\ and\ \citenamefont
  {Narang}}]{2018_Nanophotonics_7_1479_flick}%
  \BibitemOpen
  \bibfield  {author} {\bibinfo {author} {\bibfnamefont {J.}~\bibnamefont
  {Flick}}, \bibinfo {author} {\bibfnamefont {N.}~\bibnamefont {Rivera}}, \
  and\ \bibinfo {author} {\bibfnamefont {P.}~\bibnamefont {Narang}},\ }\href
  {\doibase 10.1515/nanoph-2018-0067} {\bibfield  {journal} {\bibinfo
  {journal} {Nanophotonics}\ }\textbf {\bibinfo {volume} {7}},\ \bibinfo
  {pages} {1479} (\bibinfo {year} {2018})}\BibitemShut {NoStop}%
\bibitem [{\citenamefont {Hutchison}\ \emph {et~al.}(2013)\citenamefont
  {Hutchison}, \citenamefont {Liscoi}, \citenamefont {Schwartz}, \citenamefont
  {Canaguier-Durand}, \citenamefont {Genet}, \citenamefont {Palermo},
  \citenamefont {Samori},\ and\ \citenamefont
  {Ebbesen}}]{2013_AdvMater_25_2481_hutchison}%
  \BibitemOpen
  \bibfield  {author} {\bibinfo {author} {\bibfnamefont {J.~A.}\ \bibnamefont
  {Hutchison}}, \bibinfo {author} {\bibfnamefont {A.}~\bibnamefont {Liscoi}},
  \bibinfo {author} {\bibfnamefont {T.}~\bibnamefont {Schwartz}}, \bibinfo
  {author} {\bibfnamefont {A.}~\bibnamefont {Canaguier-Durand}}, \bibinfo
  {author} {\bibfnamefont {C.}~\bibnamefont {Genet}}, \bibinfo {author}
  {\bibfnamefont {V.}~\bibnamefont {Palermo}}, \bibinfo {author} {\bibfnamefont
  {P.}~\bibnamefont {Samori}}, \ and\ \bibinfo {author} {\bibfnamefont {T.~W.}\
  \bibnamefont {Ebbesen}},\ }\href {\doibase 10.1002/adma.201203682} {\bibfield
   {journal} {\bibinfo  {journal} {Adv. Mater.}\ }\textbf {\bibinfo {volume}
  {25}},\ \bibinfo {pages} {2481} (\bibinfo {year} {2013})}\BibitemShut
  {NoStop}%
\bibitem [{\citenamefont {Orgiu}\ \emph {et~al.}(2015)\citenamefont {Orgiu},
  \citenamefont {George}, \citenamefont {Hutchison}, \citenamefont {Devaux},
  \citenamefont {Dayen}, \citenamefont {Doudin}, \citenamefont {Stellacci},
  \citenamefont {Genet}, \citenamefont {Schachenmayer}, \citenamefont {Genes},
  \citenamefont {Pupillo}, \citenamefont {Samori},\ and\ \citenamefont
  {Ebbesen}}]{2015_NatureMater_14_1123_orgiu}%
  \BibitemOpen
  \bibfield  {author} {\bibinfo {author} {\bibfnamefont {E.}~\bibnamefont
  {Orgiu}}, \bibinfo {author} {\bibfnamefont {J.}~\bibnamefont {George}},
  \bibinfo {author} {\bibfnamefont {J.~A.}\ \bibnamefont {Hutchison}}, \bibinfo
  {author} {\bibfnamefont {E.}~\bibnamefont {Devaux}}, \bibinfo {author}
  {\bibfnamefont {J.~F.}\ \bibnamefont {Dayen}}, \bibinfo {author}
  {\bibfnamefont {B.}~\bibnamefont {Doudin}}, \bibinfo {author} {\bibfnamefont
  {F.}~\bibnamefont {Stellacci}}, \bibinfo {author} {\bibfnamefont
  {C.}~\bibnamefont {Genet}}, \bibinfo {author} {\bibfnamefont
  {J.}~\bibnamefont {Schachenmayer}}, \bibinfo {author} {\bibfnamefont
  {C.}~\bibnamefont {Genes}}, \bibinfo {author} {\bibfnamefont
  {G.}~\bibnamefont {Pupillo}}, \bibinfo {author} {\bibfnamefont
  {P.}~\bibnamefont {Samori}}, \ and\ \bibinfo {author} {\bibfnamefont {T.~W.}\
  \bibnamefont {Ebbesen}},\ }\href {\doibase 10.1038/nmat4392} {\bibfield
  {journal} {\bibinfo  {journal} {Nat. Mater.}\ }\textbf {\bibinfo {volume}
  {14}},\ \bibinfo {pages} {1123} (\bibinfo {year} {2015})}\BibitemShut
  {NoStop}%
\bibitem [{\citenamefont {Thomas}\ \emph {et~al.}(2016)\citenamefont {Thomas},
  \citenamefont {George}, \citenamefont {Shalabney}, \citenamefont {Dryzhakov},
  \citenamefont {Varma}, \citenamefont {Moran}, \citenamefont {Chervy},
  \citenamefont {Zhong}, \citenamefont {Devaux}, \citenamefont {Genet},
  \citenamefont {Hutchison},\ and\ \citenamefont
  {Ebbesen}}]{2016_AngChemIE_55_11462_thomas}%
  \BibitemOpen
  \bibfield  {author} {\bibinfo {author} {\bibfnamefont {A.}~\bibnamefont
  {Thomas}}, \bibinfo {author} {\bibfnamefont {J.}~\bibnamefont {George}},
  \bibinfo {author} {\bibfnamefont {A.}~\bibnamefont {Shalabney}}, \bibinfo
  {author} {\bibfnamefont {M.}~\bibnamefont {Dryzhakov}}, \bibinfo {author}
  {\bibfnamefont {S.~J.}\ \bibnamefont {Varma}}, \bibinfo {author}
  {\bibfnamefont {J.}~\bibnamefont {Moran}}, \bibinfo {author} {\bibfnamefont
  {T.}~\bibnamefont {Chervy}}, \bibinfo {author} {\bibfnamefont
  {X.}~\bibnamefont {Zhong}}, \bibinfo {author} {\bibfnamefont
  {E.}~\bibnamefont {Devaux}}, \bibinfo {author} {\bibfnamefont
  {C.}~\bibnamefont {Genet}}, \bibinfo {author} {\bibfnamefont {J.~A.}\
  \bibnamefont {Hutchison}}, \ and\ \bibinfo {author} {\bibfnamefont {T.~W.}\
  \bibnamefont {Ebbesen}},\ }\href {\doibase 10.1002/anie.201605504} {\bibfield
   {journal} {\bibinfo  {journal} {Angew. Chem. Int. Ed.}\ }\textbf {\bibinfo
  {volume} {55}},\ \bibinfo {pages} {11462} (\bibinfo {year}
  {2016})}\BibitemShut {NoStop}%
\bibitem [{\citenamefont {Yabana}\ and\ \citenamefont
  {Bertsch}(1996)}]{1996_PRB_54_4484_yabana}%
  \BibitemOpen
  \bibfield  {author} {\bibinfo {author} {\bibfnamefont {K.}~\bibnamefont
  {Yabana}}\ and\ \bibinfo {author} {\bibfnamefont {G.~F.}\ \bibnamefont
  {Bertsch}},\ }\href {\doibase 10.1103/PhysRevB.54.4484} {\bibfield  {journal}
  {\bibinfo  {journal} {Phys. Rev. B}\ }\textbf {\bibinfo {volume} {54}},\
  \bibinfo {pages} {4484} (\bibinfo {year} {1996})}\BibitemShut {NoStop}%
\bibitem [{\citenamefont {Kuisma}\ \emph {et~al.}(2015)\citenamefont {Kuisma},
  \citenamefont {Sakko}, \citenamefont {Rossi}, \citenamefont {Larsen},
  \citenamefont {Enkovaara}, \citenamefont {Lehtovaara},\ and\ \citenamefont
  {Rantala}}]{2015_PRB_91_115431_kuisma}%
  \BibitemOpen
  \bibfield  {author} {\bibinfo {author} {\bibfnamefont {M.}~\bibnamefont
  {Kuisma}}, \bibinfo {author} {\bibfnamefont {A.}~\bibnamefont {Sakko}},
  \bibinfo {author} {\bibfnamefont {T.~P.}\ \bibnamefont {Rossi}}, \bibinfo
  {author} {\bibfnamefont {A.~H.}\ \bibnamefont {Larsen}}, \bibinfo {author}
  {\bibfnamefont {J.}~\bibnamefont {Enkovaara}}, \bibinfo {author}
  {\bibfnamefont {L.}~\bibnamefont {Lehtovaara}}, \ and\ \bibinfo {author}
  {\bibfnamefont {T.~T.}\ \bibnamefont {Rantala}},\ }\href {\doibase
  10.1103/PhysRevB.91.115431} {\bibfield  {journal} {\bibinfo  {journal} {Phys.
  Rev. B}\ }\textbf {\bibinfo {volume} {91}},\ \bibinfo {pages} {115431}
  (\bibinfo {year} {2015})}\BibitemShut {NoStop}%
\bibitem [{\citenamefont {Larsen}\ \emph {et~al.}(2009)\citenamefont {Larsen},
  \citenamefont {Vanin}, \citenamefont {Mortensen}, \citenamefont {Thygesen},\
  and\ \citenamefont {Jacobsen}}]{2009_PRB_80_195112_larsen}%
  \BibitemOpen
  \bibfield  {author} {\bibinfo {author} {\bibfnamefont {A.~H.}\ \bibnamefont
  {Larsen}}, \bibinfo {author} {\bibfnamefont {M.}~\bibnamefont {Vanin}},
  \bibinfo {author} {\bibfnamefont {J.~J.}\ \bibnamefont {Mortensen}}, \bibinfo
  {author} {\bibfnamefont {K.~S.}\ \bibnamefont {Thygesen}}, \ and\ \bibinfo
  {author} {\bibfnamefont {K.~W.}\ \bibnamefont {Jacobsen}},\ }\href {\doibase
  10.1103/PhysRevB.80.195112} {\bibfield  {journal} {\bibinfo  {journal} {Phys.
  Rev. B}\ }\textbf {\bibinfo {volume} {80}},\ \bibinfo {pages} {195112}
  (\bibinfo {year} {2009})}\BibitemShut {NoStop}%
\bibitem [{\citenamefont {Mortensen}\ \emph {et~al.}(2005)\citenamefont
  {Mortensen}, \citenamefont {Hansen},\ and\ \citenamefont
  {Jacobsen}}]{2005_PRB_71_035109_mortensen}%
  \BibitemOpen
  \bibfield  {author} {\bibinfo {author} {\bibfnamefont {J.~J.}\ \bibnamefont
  {Mortensen}}, \bibinfo {author} {\bibfnamefont {L.~B.}\ \bibnamefont
  {Hansen}}, \ and\ \bibinfo {author} {\bibfnamefont {K.~W.}\ \bibnamefont
  {Jacobsen}},\ }\href {\doibase 10.1103/PhysRevB.71.035109} {\bibfield
  {journal} {\bibinfo  {journal} {Phys. Rev. B}\ }\textbf {\bibinfo {volume}
  {71}},\ \bibinfo {pages} {035109} (\bibinfo {year} {2005})}\BibitemShut
  {NoStop}%
\bibitem [{\citenamefont {Enkovaara}\ \emph {et~al.}(2010)\citenamefont
  {Enkovaara}, \citenamefont {Rostgaard}, \citenamefont {Mortensen},
  \citenamefont {Chen}, \citenamefont {Du\l{}ak}, \citenamefont {Ferrighi},
  \citenamefont {Gavnholt}, \citenamefont {Glinsvad}, \citenamefont {Haikola},
  \citenamefont {Hansen}, \citenamefont {Kristoffersen}, \citenamefont
  {Kuisma}, \citenamefont {Larsen}, \citenamefont {Lehtovaara}, \citenamefont
  {Ljungberg}, \citenamefont {Lopez-Acevedo}, \citenamefont {Moses},
  \citenamefont {Ojanen}, \citenamefont {Olsen}, \citenamefont {Petzold},
  \citenamefont {Romero}, \citenamefont {Stausholm-M{\o}ller}, \citenamefont
  {Strange}, \citenamefont {Tritsaris}, \citenamefont {Vanin}, \citenamefont
  {Walter}, \citenamefont {Hammer}, \citenamefont {H{\"a}kkinen}, \citenamefont
  {Madsen}, \citenamefont {Nieminen}, \citenamefont {N{\o}rskov}, \citenamefont
  {Puska}, \citenamefont {Rantala}, \citenamefont {Schi{\o}tz}, \citenamefont
  {Thygesen},\ and\ \citenamefont
  {Jacobsen}}]{2010_JPhysCondMat_22_253202_enkovaara}%
  \BibitemOpen
  \bibfield  {author} {\bibinfo {author} {\bibfnamefont {J.}~\bibnamefont
  {Enkovaara}}, \bibinfo {author} {\bibfnamefont {C.}~\bibnamefont
  {Rostgaard}}, \bibinfo {author} {\bibfnamefont {J.~J.}\ \bibnamefont
  {Mortensen}}, \bibinfo {author} {\bibfnamefont {J.}~\bibnamefont {Chen}},
  \bibinfo {author} {\bibfnamefont {M.}~\bibnamefont {Du\l{}ak}}, \bibinfo
  {author} {\bibfnamefont {L.}~\bibnamefont {Ferrighi}}, \bibinfo {author}
  {\bibfnamefont {J.}~\bibnamefont {Gavnholt}}, \bibinfo {author}
  {\bibfnamefont {C.}~\bibnamefont {Glinsvad}}, \bibinfo {author}
  {\bibfnamefont {V.}~\bibnamefont {Haikola}}, \bibinfo {author} {\bibfnamefont
  {H.~A.}\ \bibnamefont {Hansen}}, \bibinfo {author} {\bibfnamefont {H.~H.}\
  \bibnamefont {Kristoffersen}}, \bibinfo {author} {\bibfnamefont
  {M.}~\bibnamefont {Kuisma}}, \bibinfo {author} {\bibfnamefont {A.~H.}\
  \bibnamefont {Larsen}}, \bibinfo {author} {\bibfnamefont {L.}~\bibnamefont
  {Lehtovaara}}, \bibinfo {author} {\bibfnamefont {M.}~\bibnamefont
  {Ljungberg}}, \bibinfo {author} {\bibfnamefont {O.}~\bibnamefont
  {Lopez-Acevedo}}, \bibinfo {author} {\bibfnamefont {P.~G.}\ \bibnamefont
  {Moses}}, \bibinfo {author} {\bibfnamefont {J.}~\bibnamefont {Ojanen}},
  \bibinfo {author} {\bibfnamefont {T.}~\bibnamefont {Olsen}}, \bibinfo
  {author} {\bibfnamefont {V.}~\bibnamefont {Petzold}}, \bibinfo {author}
  {\bibfnamefont {N.~A.}\ \bibnamefont {Romero}}, \bibinfo {author}
  {\bibfnamefont {J.}~\bibnamefont {Stausholm-M{\o}ller}}, \bibinfo {author}
  {\bibfnamefont {M.}~\bibnamefont {Strange}}, \bibinfo {author} {\bibfnamefont
  {G.~A.}\ \bibnamefont {Tritsaris}}, \bibinfo {author} {\bibfnamefont
  {M.}~\bibnamefont {Vanin}}, \bibinfo {author} {\bibfnamefont
  {M.}~\bibnamefont {Walter}}, \bibinfo {author} {\bibfnamefont
  {B.}~\bibnamefont {Hammer}}, \bibinfo {author} {\bibfnamefont
  {H.}~\bibnamefont {H{\"a}kkinen}}, \bibinfo {author} {\bibfnamefont
  {G.~K.~H.}\ \bibnamefont {Madsen}}, \bibinfo {author} {\bibfnamefont {R.~M.}\
  \bibnamefont {Nieminen}}, \bibinfo {author} {\bibfnamefont {J.~K.}\
  \bibnamefont {N{\o}rskov}}, \bibinfo {author} {\bibfnamefont
  {M.}~\bibnamefont {Puska}}, \bibinfo {author} {\bibfnamefont {T.~T.}\
  \bibnamefont {Rantala}}, \bibinfo {author} {\bibfnamefont {J.}~\bibnamefont
  {Schi{\o}tz}}, \bibinfo {author} {\bibfnamefont {K.~S.}\ \bibnamefont
  {Thygesen}}, \ and\ \bibinfo {author} {\bibfnamefont {K.~W.}\ \bibnamefont
  {Jacobsen}},\ }\href {\doibase 10.1088/0953-8984/22/25/253202} {\bibfield
  {journal} {\bibinfo  {journal} {J. Phys.: Condens. Matter}\ }\textbf
  {\bibinfo {volume} {22}},\ \bibinfo {pages} {253202} (\bibinfo {year}
  {2010})}\BibitemShut {NoStop}%
\bibitem [{\citenamefont {Rossi}\ \emph {et~al.}(2017)\citenamefont {Rossi},
  \citenamefont {Kuisma}, \citenamefont {Puska}, \citenamefont {Nieminen},\
  and\ \citenamefont {Erhart}}]{2017_JCTC_13_4779_rossi}%
  \BibitemOpen
  \bibfield  {author} {\bibinfo {author} {\bibfnamefont {T.~P.}\ \bibnamefont
  {Rossi}}, \bibinfo {author} {\bibfnamefont {M.}~\bibnamefont {Kuisma}},
  \bibinfo {author} {\bibfnamefont {M.~J.}\ \bibnamefont {Puska}}, \bibinfo
  {author} {\bibfnamefont {R.~M.}\ \bibnamefont {Nieminen}}, \ and\ \bibinfo
  {author} {\bibfnamefont {P.}~\bibnamefont {Erhart}},\ }\href {\doibase
  10.1021/acs.jctc.7b00589} {\bibfield  {journal} {\bibinfo  {journal} {J.
  Chem. Theory Comput.}\ }\textbf {\bibinfo {volume} {13}},\ \bibinfo {pages}
  {4779} (\bibinfo {year} {2017})}\BibitemShut {NoStop}%
\bibitem [{\citenamefont {Malola}\ \emph {et~al.}(2013)\citenamefont {Malola},
  \citenamefont {Lehtovaara}, \citenamefont {Enkovaara},\ and\ \citenamefont
  {H\"akkinen}}]{2013_ACSNano_7_10263_malola}%
  \BibitemOpen
  \bibfield  {author} {\bibinfo {author} {\bibfnamefont {S.}~\bibnamefont
  {Malola}}, \bibinfo {author} {\bibfnamefont {L.}~\bibnamefont {Lehtovaara}},
  \bibinfo {author} {\bibfnamefont {J.}~\bibnamefont {Enkovaara}}, \ and\
  \bibinfo {author} {\bibfnamefont {H.}~\bibnamefont {H\"akkinen}},\ }\href
  {\doibase 10.1021/nn4046634} {\bibfield  {journal} {\bibinfo  {journal} {ACS
  Nano}\ }\textbf {\bibinfo {volume} {7}},\ \bibinfo {pages} {10263} (\bibinfo
  {year} {2013})}\BibitemShut {NoStop}%
\bibitem [{\citenamefont {Knight}\ \emph {et~al.}(2014)\citenamefont {Knight},
  \citenamefont {King}, \citenamefont {Liu}, \citenamefont {Everitt},
  \citenamefont {Nordlander},\ and\ \citenamefont
  {Nalas}}]{2014_ACSNano_8_834_knight}%
  \BibitemOpen
  \bibfield  {author} {\bibinfo {author} {\bibfnamefont {M.~W.}\ \bibnamefont
  {Knight}}, \bibinfo {author} {\bibfnamefont {N.~S.}\ \bibnamefont {King}},
  \bibinfo {author} {\bibfnamefont {L.}~\bibnamefont {Liu}}, \bibinfo {author}
  {\bibfnamefont {H.~O.}\ \bibnamefont {Everitt}}, \bibinfo {author}
  {\bibfnamefont {P.}~\bibnamefont {Nordlander}}, \ and\ \bibinfo {author}
  {\bibfnamefont {N.~J.}\ \bibnamefont {Nalas}},\ }\href {\doibase
  10.1021/nn405495q} {\bibfield  {journal} {\bibinfo  {journal} {ACS Nano}\
  }\textbf {\bibinfo {volume} {8}},\ \bibinfo {pages} {834} (\bibinfo {year}
  {2014})}\BibitemShut {NoStop}%
\bibitem [{\citenamefont {Koenderink}(2010)}]{OL_35_4208_koenderink}%
  \BibitemOpen
  \bibfield  {author} {\bibinfo {author} {\bibfnamefont {A.~F.}\ \bibnamefont
  {Koenderink}},\ }\href {\doibase 10.1364/OL.35.004208} {\bibfield  {journal}
  {\bibinfo  {journal} {Opt. Lett.}\ }\textbf {\bibinfo {volume} {35}},\
  \bibinfo {pages} {4208} (\bibinfo {year} {2010})}\BibitemShut {NoStop}%
\bibitem [{\citenamefont {Wu}\ \emph {et~al.}(2010)\citenamefont {Wu},
  \citenamefont {Gray},\ and\ \citenamefont {Pelton}}]{OE_18_23633_wu}%
  \BibitemOpen
  \bibfield  {author} {\bibinfo {author} {\bibfnamefont {X.}~\bibnamefont
  {Wu}}, \bibinfo {author} {\bibfnamefont {S.~K.}\ \bibnamefont {Gray}}, \ and\
  \bibinfo {author} {\bibfnamefont {M.}~\bibnamefont {Pelton}},\ }\href
  {\doibase 10.1364/OE.18.023633} {\bibfield  {journal} {\bibinfo  {journal}
  {Opt. Express}\ }\textbf {\bibinfo {volume} {18}},\ \bibinfo {pages} {23633}
  (\bibinfo {year} {2010})}\BibitemShut {NoStop}%
\bibitem [{\citenamefont {Yang}\ \emph {et~al.}(2016)\citenamefont {Yang},
  \citenamefont {Antosiewicz},\ and\ \citenamefont
  {Shegai}}]{2016_OpEx_24_20373_yang}%
  \BibitemOpen
  \bibfield  {author} {\bibinfo {author} {\bibfnamefont {Z.-J.}\ \bibnamefont
  {Yang}}, \bibinfo {author} {\bibfnamefont {T.~J.}\ \bibnamefont
  {Antosiewicz}}, \ and\ \bibinfo {author} {\bibfnamefont {T.}~\bibnamefont
  {Shegai}},\ }\href {\doibase 10.1364/OE.24.020373} {\bibfield  {journal}
  {\bibinfo  {journal} {Opt. Express}\ }\textbf {\bibinfo {volume} {24}},\
  \bibinfo {pages} {20373} (\bibinfo {year} {2016})}\BibitemShut {NoStop}%
\bibitem [{\citenamefont {Benz}\ \emph {et~al.}(2016)\citenamefont {Benz},
  \citenamefont {Schmidt}, \citenamefont {Dreismann}, \citenamefont
  {Chikkaraddy}, \citenamefont {Zhang}, \citenamefont {Demetriadou},
  \citenamefont {Carnegie}, \citenamefont {Ohadi}, \citenamefont {de~Nijs},
  \citenamefont {Esteban}, \citenamefont {Aizpurua},\ and\ \citenamefont
  {Baumberg}}]{2016_Sci_354_726_benz}%
  \BibitemOpen
  \bibfield  {author} {\bibinfo {author} {\bibfnamefont {F.}~\bibnamefont
  {Benz}}, \bibinfo {author} {\bibfnamefont {M.~K.}\ \bibnamefont {Schmidt}},
  \bibinfo {author} {\bibfnamefont {A.}~\bibnamefont {Dreismann}}, \bibinfo
  {author} {\bibfnamefont {R.}~\bibnamefont {Chikkaraddy}}, \bibinfo {author}
  {\bibfnamefont {Y.}~\bibnamefont {Zhang}}, \bibinfo {author} {\bibfnamefont
  {A.}~\bibnamefont {Demetriadou}}, \bibinfo {author} {\bibfnamefont
  {C.}~\bibnamefont {Carnegie}}, \bibinfo {author} {\bibfnamefont
  {H.}~\bibnamefont {Ohadi}}, \bibinfo {author} {\bibfnamefont
  {B.}~\bibnamefont {de~Nijs}}, \bibinfo {author} {\bibfnamefont
  {R.}~\bibnamefont {Esteban}}, \bibinfo {author} {\bibfnamefont
  {J.}~\bibnamefont {Aizpurua}}, \ and\ \bibinfo {author} {\bibfnamefont
  {J.~J.}\ \bibnamefont {Baumberg}},\ }\href {\doibase 10.1126/science.aah5243}
  {\bibfield  {journal} {\bibinfo  {journal} {Science}\ }\textbf {\bibinfo
  {volume} {354}},\ \bibinfo {pages} {726} (\bibinfo {year}
  {2016})}\BibitemShut {NoStop}%
\bibitem [{\citenamefont {Antosiewicz}\ \emph {et~al.}(2012)\citenamefont
  {Antosiewicz}, \citenamefont {Apell}, \citenamefont {Claudio},\ and\
  \citenamefont {K\"all}}]{OE_20_524_tja}%
  \BibitemOpen
  \bibfield  {author} {\bibinfo {author} {\bibfnamefont {T.~J.}\ \bibnamefont
  {Antosiewicz}}, \bibinfo {author} {\bibfnamefont {S.~P.}\ \bibnamefont
  {Apell}}, \bibinfo {author} {\bibfnamefont {V.}~\bibnamefont {Claudio}}, \
  and\ \bibinfo {author} {\bibfnamefont {M.}~\bibnamefont {K\"all}},\ }\href
  {\doibase 10.1364/OE.20.000524} {\bibfield  {journal} {\bibinfo  {journal}
  {Opt. Express}\ }\textbf {\bibinfo {volume} {20}},\ \bibinfo {pages} {524}
  (\bibinfo {year} {2012})}\BibitemShut {NoStop}%
\bibitem [{\citenamefont {Antosiewicz}\ \emph {et~al.}(2014)\citenamefont
  {Antosiewicz}, \citenamefont {Apell},\ and\ \citenamefont
  {Shegai}}]{ACSPhoton_1_454_tja}%
  \BibitemOpen
  \bibfield  {author} {\bibinfo {author} {\bibfnamefont {T.~J.}\ \bibnamefont
  {Antosiewicz}}, \bibinfo {author} {\bibfnamefont {S.~P.}\ \bibnamefont
  {Apell}}, \ and\ \bibinfo {author} {\bibfnamefont {T.}~\bibnamefont
  {Shegai}},\ }\href {\doibase 10.1021/ph500032d} {\bibfield  {journal}
  {\bibinfo  {journal} {ACS Photonics}\ }\textbf {\bibinfo {volume} {1}},\
  \bibinfo {pages} {454} (\bibinfo {year} {2014})}\BibitemShut {NoStop}%
\bibitem [{\citenamefont {Zhang}\ \emph {et~al.}(2015)\citenamefont {Zhang},
  \citenamefont {Chen},\ and\ \citenamefont {Li}}]{JPCC_119_11858_zhang}%
  \BibitemOpen
  \bibfield  {author} {\bibinfo {author} {\bibfnamefont {C.}~\bibnamefont
  {Zhang}}, \bibinfo {author} {\bibfnamefont {B.-Q.}\ \bibnamefont {Chen}}, \
  and\ \bibinfo {author} {\bibfnamefont {Z.-Y.}\ \bibnamefont {Li}},\ }\href
  {\doibase 10.1021/acs.jpcc.5b02653} {\bibfield  {journal} {\bibinfo
  {journal} {J. Phys. Chem. C}\ }\textbf {\bibinfo {volume} {119}},\ \bibinfo
  {pages} {11858} (\bibinfo {year} {2015})}\BibitemShut {NoStop}%
\bibitem [{\citenamefont {Chikkaraddy}\ \emph {et~al.}(2016)\citenamefont
  {Chikkaraddy}, \citenamefont {de~Nijs}, \citenamefont {Benz}, \citenamefont
  {Barrow}, \citenamefont {Scherman}, \citenamefont {Rosta}, \citenamefont
  {Demetriadou}, \citenamefont {Fox}, \citenamefont {Hess},\ and\ \citenamefont
  {Baumberg}}]{BaumbergNature2016}%
  \BibitemOpen
  \bibfield  {author} {\bibinfo {author} {\bibfnamefont {R.}~\bibnamefont
  {Chikkaraddy}}, \bibinfo {author} {\bibfnamefont {B.}~\bibnamefont
  {de~Nijs}}, \bibinfo {author} {\bibfnamefont {F.}~\bibnamefont {Benz}},
  \bibinfo {author} {\bibfnamefont {S.~J.}\ \bibnamefont {Barrow}}, \bibinfo
  {author} {\bibfnamefont {O.~A.}\ \bibnamefont {Scherman}}, \bibinfo {author}
  {\bibfnamefont {E.}~\bibnamefont {Rosta}}, \bibinfo {author} {\bibfnamefont
  {A.}~\bibnamefont {Demetriadou}}, \bibinfo {author} {\bibfnamefont
  {P.}~\bibnamefont {Fox}}, \bibinfo {author} {\bibfnamefont {O.}~\bibnamefont
  {Hess}}, \ and\ \bibinfo {author} {\bibfnamefont {J.~J.}\ \bibnamefont
  {Baumberg}},\ }\href {\doibase 10.1038/nature17974} {\bibfield  {journal}
  {\bibinfo  {journal} {Nature}\ }\textbf {\bibinfo {volume} {535}},\ \bibinfo
  {pages} {127} (\bibinfo {year} {2016})}\BibitemShut {NoStop}%
\bibitem [{\citenamefont {George}\ \emph {et~al.}(2015)\citenamefont {George},
  \citenamefont {Wang}, \citenamefont {Chervy}, \citenamefont
  {Canaguier-Durand}, \citenamefont {Schaeffer}, \citenamefont {Lehn},
  \citenamefont {Hutchison}, \citenamefont {Genet},\ and\ \citenamefont
  {Ebbesen}}]{2015_FaradDisc_178_281_george}%
  \BibitemOpen
  \bibfield  {author} {\bibinfo {author} {\bibfnamefont {J.}~\bibnamefont
  {George}}, \bibinfo {author} {\bibfnamefont {S.}~\bibnamefont {Wang}},
  \bibinfo {author} {\bibfnamefont {T.}~\bibnamefont {Chervy}}, \bibinfo
  {author} {\bibfnamefont {A.}~\bibnamefont {Canaguier-Durand}}, \bibinfo
  {author} {\bibfnamefont {G.}~\bibnamefont {Schaeffer}}, \bibinfo {author}
  {\bibfnamefont {J.-M.}\ \bibnamefont {Lehn}}, \bibinfo {author}
  {\bibfnamefont {J.~A.}\ \bibnamefont {Hutchison}}, \bibinfo {author}
  {\bibfnamefont {C.}~\bibnamefont {Genet}}, \ and\ \bibinfo {author}
  {\bibfnamefont {T.~W.}\ \bibnamefont {Ebbesen}},\ }\href {\doibase
  10.1039/C4FD00197D} {\bibfield  {journal} {\bibinfo  {journal} {Faraday
  Discuss.}\ }\textbf {\bibinfo {volume} {178}},\ \bibinfo {pages} {281}
  (\bibinfo {year} {2015})}\BibitemShut {NoStop}%
\bibitem [{\citenamefont {Perdew}\ \emph {et~al.}(1996)\citenamefont {Perdew},
  \citenamefont {Burke},\ and\ \citenamefont
  {Ernzerhof}}]{1996_PRL_77_3865_perdew}%
  \BibitemOpen
  \bibfield  {author} {\bibinfo {author} {\bibfnamefont {J.~P.}\ \bibnamefont
  {Perdew}}, \bibinfo {author} {\bibfnamefont {K.}~\bibnamefont {Burke}}, \
  and\ \bibinfo {author} {\bibfnamefont {M.}~\bibnamefont {Ernzerhof}},\ }\href
  {\doibase 10.1103/PhysRevLett.77.3865} {\bibfield  {journal} {\bibinfo
  {journal} {Phys. Rev. Lett.}\ }\textbf {\bibinfo {volume} {77}},\ \bibinfo
  {pages} {3865} (\bibinfo {year} {1996})}\BibitemShut {NoStop}%
\bibitem [{\citenamefont {Perdew}\ \emph {et~al.}(1997)\citenamefont {Perdew},
  \citenamefont {Burke},\ and\ \citenamefont
  {Ernzerhof}}]{1997_PRL_78_1396_perdew}%
  \BibitemOpen
  \bibfield  {author} {\bibinfo {author} {\bibfnamefont {J.~P.}\ \bibnamefont
  {Perdew}}, \bibinfo {author} {\bibfnamefont {K.}~\bibnamefont {Burke}}, \
  and\ \bibinfo {author} {\bibfnamefont {M.}~\bibnamefont {Ernzerhof}},\ }\href
  {\doibase 10.1103/PhysRevLett.78.1396} {\bibfield  {journal} {\bibinfo
  {journal} {Phys. Rev. Lett.}\ }\textbf {\bibinfo {volume} {78}},\ \bibinfo
  {pages} {1396} (\bibinfo {year} {1997})}\BibitemShut {NoStop}%
\bibitem [{\citenamefont {Bl\"ochl}(1994)}]{1994_PRB_50_17953_blochl}%
  \BibitemOpen
  \bibfield  {author} {\bibinfo {author} {\bibfnamefont {P.~E.}\ \bibnamefont
  {Bl\"ochl}},\ }\href {\doibase 10.1103/PhysRevB.50.17953} {\bibfield
  {journal} {\bibinfo  {journal} {Phys. Rev. B}\ }\textbf {\bibinfo {volume}
  {50}},\ \bibinfo {pages} {17953} (\bibinfo {year} {1994})}\BibitemShut
  {NoStop}%
\bibitem [{\citenamefont {Rossi}\ \emph {et~al.}(2015)\citenamefont {Rossi},
  \citenamefont {Lehtola}, \citenamefont {Sakko}, \citenamefont {Puska},\ and\
  \citenamefont {Nieminen}}]{2015_JChemPhys_142_094114_rossi}%
  \BibitemOpen
  \bibfield  {author} {\bibinfo {author} {\bibfnamefont {T.~P.}\ \bibnamefont
  {Rossi}}, \bibinfo {author} {\bibfnamefont {S.}~\bibnamefont {Lehtola}},
  \bibinfo {author} {\bibfnamefont {A.}~\bibnamefont {Sakko}}, \bibinfo
  {author} {\bibfnamefont {M.~J.}\ \bibnamefont {Puska}}, \ and\ \bibinfo
  {author} {\bibfnamefont {R.~M.}\ \bibnamefont {Nieminen}},\ }\href {\doibase
  10.1063/1.4913739} {\bibfield  {journal} {\bibinfo  {journal} {J. Chem.
  Phys.}\ }\textbf {\bibinfo {volume} {142}},\ \bibinfo {pages} {094114}
  (\bibinfo {year} {2015})}\BibitemShut {NoStop}%
\end{thebibliography}%

\end{document}